\documentclass[%
reprint,
 amsmath,amssymb,
 aps,
]{revtex4-2}

\usepackage{graphicx}
\usepackage{dcolumn}
\usepackage{bm}
\usepackage[colorlinks=true,
            linkcolor=blue,
            urlcolor=blue,
            citecolor=blue]{hyperref}
\usepackage{svg}
\usepackage{amsmath}

\definecolor{myorange}{RGB}{255,102,51}

\def\fullnameAllCap{Parameter Inference from a Non-stationary Unknown Process}

\def\problemname{PINUP}
\def\fullmethodname{feature-based Slow Feature Analysis}
\def\ourmethod{$f$-SFA}
\def\baseline{$f$-SFA-mean-var}

\begin{document}

\preprint{APS/123-QED}

\title{Parameter inference from a non-stationary unknown process using statistical feature-based slow feature analysis}

\author{Kieran S. Owens}
\email{kowe6911@uni.sydney.edu.au}
\affiliation{%
School of Physics, The University of Sydney, Camperdown 2006, NSW Australia.
}%
\author{Masako Tamaki}
\affiliation{%
RIKEN Center for Brain Science, Saitama 351-0198, Japan.
}%
\affiliation{%
RIKEN Pioneering Research Institute, Saitama 351-0198, Japan.
}
\author{Ben D. Fulcher}%
\email{ben.fulcher@sydney.edu.au}
\affiliation{%
School of Physics, The University of Sydney, Camperdown 2006, NSW Australia.
}%

\date{\today}

\begin{abstract}
Non-stationary phenomena are ubiquitous, with examples to be found in climatological measurements, brain activity, and the behavior of financial markets.
Starting with a time series from a non-stationary process, a key challenge is to infer the time-varying parameters that underlie the non-stationarity in these systems, without requiring a generative model of the dynamics to be learned.
This problem is referred to as \fullnameAllCap{} (\problemname{}).
Here we introduce a \problemname{} method called \fullmethodname{} (\ourmethod{}) comprising the computation of time-series features across sliding windows, followed by dimension reduction using slow feature analysis (SFA).
This allows us to detect slow variation in a potentially wide range of statistical properties of the measured dynamics on a timescale determined by the window length.
Crucially, using a comprehensive time-series feature set avoids the subjectivity of feature selection, while the SFA slowness constraint overcomes the bias towards irrelevant correlated features seen with variance-based dimension reduction.
The performance of \ourmethod{} surpasses that of four benchmark \problemname{} methods across a diverse range of non-stationary chaotic processes, and we explore the impact of various parameters on performance, including observation noise, parameter timescales, parameter amplitudes, and unseen parameter values.
Further, applying \ourmethod{} to sleep polysomnography data, we show that it is able to infer a time-varying parameter underlying the non-stationary sleep recordings that closely tracks depth of sleep.
To our knowledge, this work presents the first comparative study of \problemname{} methods, and we demonstrate that \ourmethod{} is a simple, effective, and noise-robust approach for quantifying non-stationarity from time series, that can be applied in a range of fields.
\end{abstract}

\maketitle


\section{\label{sec:level1}Introduction}

Time-series analysis tools are widely used in science and industry, for example, when forecasting climate trends from meteorological data, diagnosing disease states using brain activity, or predicting the activity of financial markets.
Notably, many time-series analysis methods make the simplifying assumption that the analyzed time series are stationary.
Conventionally, \textit{strong stationarity} means that all conditional probabilities are constant in time, while \textit{weak stationarity} means that the first and second moments of the distribution of the values of a time series are constant \cite{schreiberClassificationTimeSeries1997}.
So defined, these properties apply to processes rather than finite time series, and can only be evaluated in the large-data limit.

A practical alternative definition that allows data-driven inference is to designate a process as non-stationary when one can identify one or more statistics that vary over some timescale.
For example, variation in the joint probability $p(x_t,...x_{t+W})$ of a univariate process over time $t$ (on the timescale of the window length $W$) may in turn produce variation in properties such as variance or autocorrelation that can be estimated statistically.
Given a univariate times series of window length $W$, we define a time-series statistic to be a map $f: \mathbb{R}^W \rightarrow \mathbb{R}$ from the time series to a real-valued time-series feature.
In practice, assessing such properties requires specifying a window size $W$ over which the variation is evaluated.
Inferring and tracking sources of non-stationarity could enhance our understanding of the dynamics underlying non-stationary processes that are studied in a range of fields, including spike trains in neuroscience \cite{gourevitchSimpleIndicatorNonstationarity2007}, population outbreaks in ecology \cite{rollinsonWorkingSpaceTime2021}, and extreme weather events in meteorology \cite{slaterNonstationaryWeatherWater2021}.

Here we focus on the problem of \fullnameAllCap{} (\problemname{}), as defined previously \cite{owensParameterInferenceNonstationary2024}.
Given a time series generated by some process where non-stationarity is driven by one or more time-varying parameters (TVPs), the goal of \problemname{} is to infer the TVPs using only the observed time series.
Crucially, this inference is performed without knowledge of (nor a requirement to infer) a mathematical model of the underlying process, so that standard parameter-fitting methods used for inverse problems involving known model equations are not applicable \cite{tarantolaInverseProblemTheory2005}.
We previously reviewed existing approaches to \problemname{} \cite{owensParameterInferenceNonstationary2024}, where we grouped them into six categories: (1) dimension reduction, (2) statistical time-series features, (3) recurrence quantification analysis, (4) prediction error, (5) phase space partitioning, and (6) Bayesian inference.
In addition to quantifying and studying non-stationarity, per se, two areas of potential application for \problemname{} methods include anticipating critical transitions and system identification for non-stationary processes.
Critical transitions are important phenomena in a range of systems, e.g., financial crises or climate events related to market volatility or global temperatures, respectively.
Recent research has shown that critical transitions are more easily predicted if the time series of a TVP that drives dynamical variation is also available, whether that TVP is known \cite{patelUsingMachineLearning2021a} or inferred \cite{tokudaPredictionUnobservedBifurcation2024}.
Similarly, a TVP learned via \problemname{} could be provided as an input to a system identification method for non-stationary processes, such as the sparse identification of nonlinear dynamics with control parameters (SINDyCP) algorithm \cite{nicolaouDatadrivenDiscoveryExtrapolation2023c}.

Ideally, a \problemname{} algorithm would track the variation of the joint probability distribution of a process across some timescale, but we do not expect this problem to be tractable in general for finite data of length $T$ (since, in the absence of assumptions about the form of the underlying process, this results in a non-parametric density-estimation problem , which is challenging for longer time series and/or larger numbers of variables \cite{scottFeasibilityMultivariateDensity1991, otneimNonParametricEstimationConditional2016}).
An alternative approach is to use statistical time-series features to track parameter variation, as exemplified by the work of \citet{guttlerReconstructionParameterSpaces2001}, which showed that the parameter spaces underlying certain datasets can be reconstructed using well-chosen statistics such as mean, variance, two-point correlation, and the Lyapunov exponent.
However, the key weakness of this approach is that expert knowledge is required to select the time-series statistics, making it inappropriate as a general approach to \problemname{}, where such \textit{a priori} knowledge is unavailable.
One solution to this problem of tailoring features for \problemname{} is to start with a large set of candidate statistical features and use the properties of the data to select those that are most relevant to the given problem.
This approach can be operationalized by using one of the comprehensive open-source feature sets that are now available such as \textit{hctsa} \cite{fulcherHctsaComputationalFramework2017} (which contains over $7000$ univariate features), among others \cite{hendersonEmpiricalEvaluationTimeSeries2021, cliffUnifyingPairwiseInteractions2023a, moorePyhctsaPythonPackage2026}.
Starting with a set of time-series statistics that are sensitive to a range of dynamical properties, the corresponding statistical features can be computed across sliding windows of a (possibly multivariate) time series, thereby indexing the time-variation of the local statistical properties of the process.
Next, dimension reduction can be applied to the time-series of statistical features to find the combination of features that best tracks the statistical variation in the dataset that is driven by underlying TVP(s).
This achieves a balance between a purely feature-based approach to \problemname{} (for which the subjectivity of feature choice is a problem), and a pure dimension-reduction approach (that is more data-driven but lacks the inductive biases needed to track a range of complex statistical properties).

Precedent for using a large set of time-series features to index parameter variation can be found in \citet{fulcherInterpretableModelfreeInference2026} where the combination of time-series features and dimension reduction (specifically, principal component analysis (PCA) or Isomap) was used to infer parametric variation across an \textit{unordered} collection of time series.
In contrast, in accordance with \problemname{}, we seek to track parameter variation over time for a single (though possibly multivariate) non-stationary time series.
Crucially, an issue that arises when using variance-based dimension-reduction methods such as PCA in this setting is that correlated features within a large feature set may produce spurious, dominant directions of shared variance that are not related to parameter variation, leading to failure of parameter inference \cite{harrisInferringParametricVariation2021, fulcherInterpretableModelfreeInference2026}.
Given that our objective is to resolve parameter variation over time, we can address this key weakness of variance-based dimension reduction applied to large feature sets by instead making use of one of the many time-series dimension-reduction methods that utilize temporal properties such as slowness, predictability, or autocorrelation, instead of covariance \cite{owensTimeseriesDimensionReduction}, which we hypothesize will be robust to the presence of irrelevant correlated features.
The key idea explored in this paper is to combine a dynamically rich time-series feature set (thereby overcoming the subjectivity of feature selection) with a dimension-reduction method that utilizes temporal structure (thereby overcoming the weakness faced by variance-based dimension reduction in the presence of irrelevant correlated features) to reliably estimate diverse statistical manifestations of non-stationary variation from time-series observations.

In this paper we introduce a \problemname{} method that combines a set of canonical time-series features, that are computed over sliding windows, with dimension reduction via slow feature analysis (SFA) \cite{wiskottSlowFeatureAnalysis2002}.
We call the method \fullmethodname{} (\ourmethod{}).
In principle, any time-series feature set could be used to provide a statistical signature of each window of the dynamics, but in this work we use \textit{catch24} which consists of 24 statistical features: (i) the 22 interpretable time-series features contained in \textit{catch22} (derivied from \textit{hctsa}, which provide a compact and computationally efficient summary of the interdisciplinary time-series literature \cite{lubbaCatch22CAnonicalTimeseries2019}), and (ii) 2 additional distributional moments, the mean and standard deviation.
SFA is a linear dimension-reduction method that can be applied to any multivariate signal to extract the most slowly varying components, and here we apply it to multivariate time series of statistical features.
Consequently, we expect \problemname{} via \ourmethod{} to be possible whenever one or more statistical properties that vary over time are monotonically related to a TVP that evolves on a sufficiently slow timescale relative to that of both the observed process and the observation window length.
In turn, \ourmethod{} allows us to identify the types of dynamical properties that vary under the influence of slowly varying parameters, and to make connections to the time-series theory underlying the different statistics.
Finally, although many \problemname{} methods have been proposed \cite{owensParameterInferenceNonstationary2024}, to our knowledge most papers do not feature any comparison between methods, or else conduct only a limited comparison, say, with only one other method, and this issue has been compounded by a lack of open-source software implementations of \problemname{} algorithms.
To address these gaps, we will compare the performance of \ourmethod{} to several other \problemname{} algorithms, the software implementations of which will be made publicly available as an accompaniment to this manuscript.


This paper is structured as follows.
First we detail our methods (in Sec.~\ref{sec_methods}), including a formulation of the \problemname{} problem, and a description of the \ourmethod{} algorithm and the other \problemname{} methods that were chosen for comparison.
We then present the results of several numerical experiments (in Sec.~\ref{sec_results}), including a comparison of SFA versus PCA in the presence of correlated features (Sec.~\ref{feature_feature}), a comparison between \ourmethod{} and four other \problemname{} methods (Sec.~\ref{experiment_comparative}), and an examination of how the performance of \ourmethod{} is affected by factors such as parameter timescale (Sec.~\ref{experiment_timescale}), parameter amplitude (Sec.~\ref{experiment_amplitude}), and unseen parameter values (Sec.~\ref{experiment_ts_length}).
Finally, we apply \ourmethod{} to sleep polysomnography data and demonstrate that it can infer a time-series component from data that tracks sleep depth (in Sec.~\ref{sleep_application}).

\begin{figure*}
\includegraphics[width=480pt]{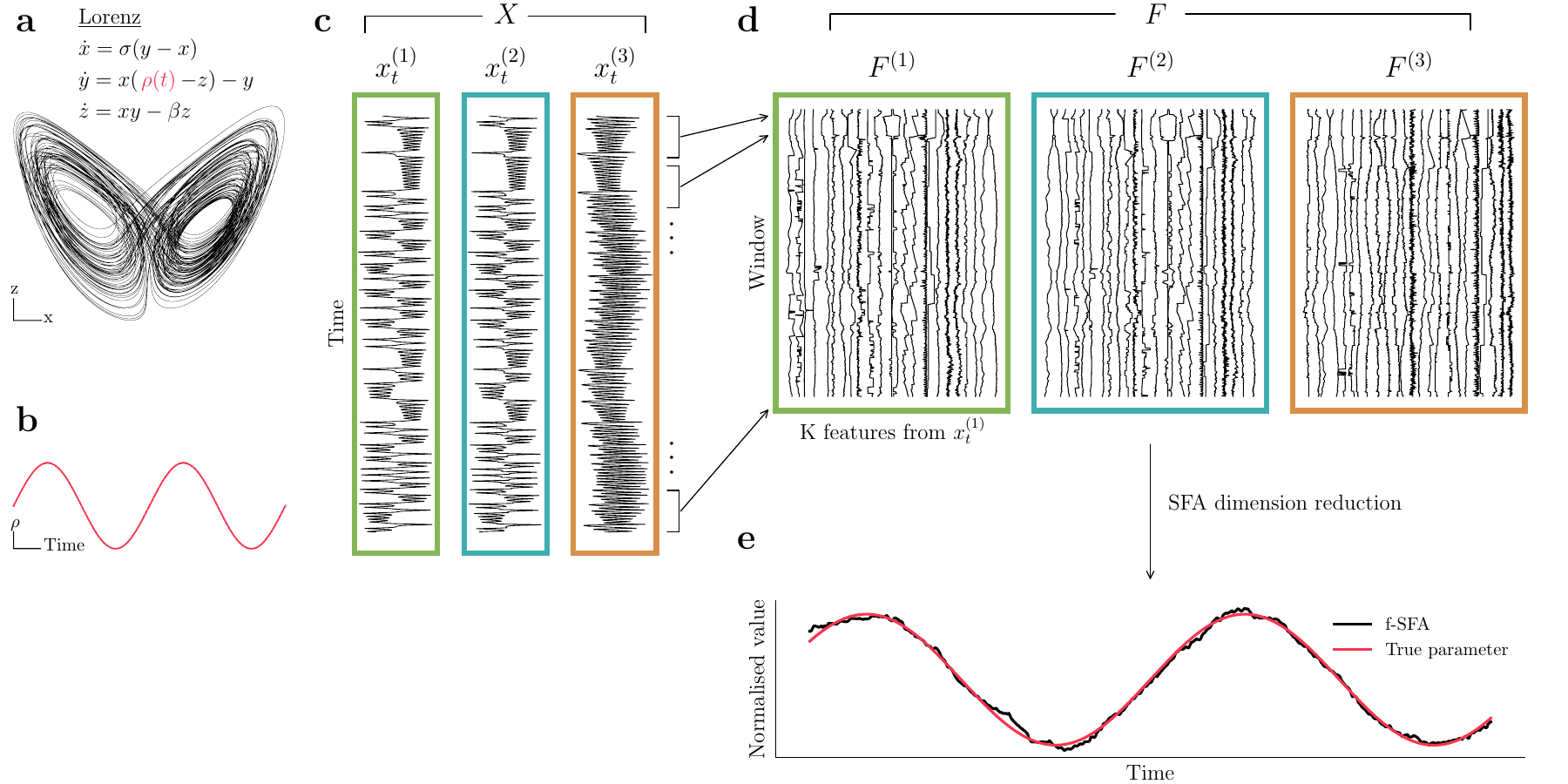}
\vspace{-0cm}
\caption{\label{fig_summary}
\textbf{Summary of the feature-based slow feature analysis (\ourmethod{}) method for \fullnameAllCap{} (\problemname{}) introduced here.}
This figure schematically introduces \ourmethod{}, an algorithm for inferring one or more time-varying parameters from an observed non-stationary time series.
(a) The Lorenz process driven by a single TVP, $\rho(t)$, was chosen as a simple canonical case study on which to demonstrate \ourmethod{}.
A 2-dimensional projection of a non-stationary Lorenz process is shown.
(b) We varied the $\rho(t)$ parameter sinusoidally, shown in red, resulting in a non-stationary time series.
The aim of \problemname{} is to infer this time variation directly from time-series data, without any other knowledge (specifically, without the model equations).
(c) In this example, three non-stationary time series $x_t^{(i)}$ are measured, the concatenation of which comprise a multivariate time series $X$.
(d) A set of $K$ sliding-window statistics is computed for each temporal window, for each observed time series to obtain a multivariate time series $F^{(i)}$ of $K$ statistical features, the concatenation of which is denoted $F$.
This example uses the \textit{catch24} feature set to provide a statistical signature of the dynamics in each window, with $K = 24$, and depicts the simplified case of non-overlapping windows.
Thus, for each $F^{(i)}$, each trace represents one of the \textit{catch24} features.
(e) Dimension reduction is performed by applying slow feature analysis (SFA) to $F$, the multivariate time series of windowed statistical features from each variable of the original time series $X$.
}
\end{figure*}


\section{\label{sec_methods}Methods}

Our method, \ourmethod{}, for feature-based parameter inference is depicted schematically in Fig.~\ref{fig_summary} to give a high-level overview of the main algorithmic steps applied to a simple, canonical case study of a non-stationary Lorenz process.
Starting with a non-stationary process [Fig.~\ref{fig_summary}(a)] driven by a TVP [$\rho(t)$, Fig.~\ref{fig_summary}(b)], a (possibly multivariate) time series $X \in \mathbb{R}^{T \times D}$ of length $T$ and dimension $D$ is observed [Fig.~\ref{fig_summary}(c)].
For each observed variable, a set of $K$ statistics is computed across sliding windows, yielding multivariate time series of statistical features [Fig.~\ref{fig_summary}(d)], reflecting the time-varying statistical properties of the underlying non-stationary process on a timescale determined by the window $W$.
Dimension reduction is then performed using SFA, which extracts a subspace capturing the direction of the slowest statistical variation \cite{wiskottSlowFeatureAnalysis2002}, and the slowest component is taken as the (unsupervised) inferred underlying TVP [Fig.~\ref{fig_summary}(e), black], which in this case successfully captures the underlying sinusoidal variation of the true parameter [from Fig.~\ref{fig_summary}(b)].
This paper mainly focuses on the case of a single TVP, but in Appendix~\ref{appendix_multiple} we provide a case study of how \ourmethod{} can be used to infer multiple time-varying sources of non-stationarity.
In what follows, we describe the \problemname{} problem (Sec.~\ref{sec_formulation}), the \ourmethod{} algorithm (Sec.~\ref{sec_fsfa}), four \problemname{} methods that were selected for comparison (Sec.~\ref{sec_other_methods}), and our procedure for simulating non-stationary processes and comparing \problemname{} methods (Sec.~\ref{sec_nonstat_procs}).

\subsection{\label{sec_formulation} \problemname{} problem formulation}

The \problemname{} problem class is summarized as follows: given an observed time-series dataset resulting from some unknown process $\mathcal{M}$ under the influence of TVP(s) $\boldsymbol{\theta}(t)$ (the discretization of which is $\boldsymbol{\theta}_t$), our goal is to infer an approximation of the TVP(s), denoted $\hat{ \boldsymbol{\theta}}_t$ \cite{owensParameterInferenceNonstationary2024}.
We consider a multivariate (in general) time series to be generated by a non-stationary process $\mathcal{M}$, the joint probability distribution of which depends on one or more time-varying parameters, $\boldsymbol{\theta}(t)$, across some time interval $t \in [0,T_{\text{max}}]$.
For example, $\mathcal{M}$ could represent a system of ODEs $\frac{d}{dt} \boldsymbol{x} = \boldsymbol{g}(\boldsymbol{x}, \boldsymbol{\theta}(t))$, or an iterative map $\boldsymbol{x}_{t+1} = \boldsymbol{G}(\boldsymbol{x}_t, \boldsymbol{\theta}_t)$, among other possibilities.
However, we do not require $\mathcal{M}$ to be closed form, or even to be able to be written down.
In practice, we obtain a time series of length $T$, which here we consider to be sampled at a uniform sampling rate, as $t = 0, \Delta t, 2\Delta t, ..., (T-1) \Delta t$.
We denote a time-series realization from $\mathcal{M}(\boldsymbol{\theta}(t))$ as $X \in \mathbb{R}^{T \times D}$, where $D$ is the number of time-series variables, and the value at time $t$ of variable $i$ is $x_t^{(i)}$.
We can write $X$ as:
\begin{equation} X =
\begin{bmatrix}
    \rule[.5ex]{1em}{0.4pt} & \boldsymbol{x}_1 & \rule[.5ex]{1em}{0.4pt}\\
    \rule[.5ex]{1em}{0.4pt} & \boldsymbol{x}_2 & \rule[.5ex]{1em}{0.4pt}\\
    & \vdots \\
    \rule[.5ex]{1em}{0.4pt} & \boldsymbol{x}_T & \rule[.5ex]{1em}{0.4pt}
\end{bmatrix}\,,
\end{equation}
where time increases down each column and the multivariate measurements at each time point, $\boldsymbol{x}_i$, span rows.


\subsection{\label{sec_fsfa}$f$-SFA}

Feature-based slow feature analysis (\ourmethod{}) is a  \problemname{} method combining statistical time-series features and dimension reduction using SFA.
As noted above, using a feature set capable of tracking a range of dynamical properties mitigates the subjectivity of feature selection, while the inductive bias of SFA towards slowness overcomes the bias towards irrelevant correlated features seen with variance-based methods like PCA.
\ourmethod{} starts with a time series with $D\geq 1$ variables and computes a matrix of statistical time-series features, reflecting the variation of statistical properties over time [Fig.~\ref{fig_summary}(c)], after which SFA is applied to obtain a subspace capturing the direction(s) of slowest variation of the measured statistical properties [Fig.~\ref{fig_summary}(d)].
The intuition underlying \ourmethod{} is that, when the TVP(s) affect the statistics joint distribution of the underlying process, then this variation can be detected using statistics that are sensitive to that change.
Moreover, when the TVP(s) vary slowly relative to the observed dynamics, SFA can be applied to the matrix of time-varying statistics to infer components that are candidates for tracking the underlying TVP(s).

The implementation of \ourmethod{} used throughout this work, that combines \textit{catch24} feature extraction \cite{lubbaCatch22CAnonicalTimeseries2019, lubbaDynamicsAndNeuralSystemsCatch22V02022} with SFA using \textit{sklearn-sfa} \cite{WiskottlabSklearnsfa2024}, is available in the accompanying GitHub repository: \url{https://github.com/KieranOwens/fsfa}.


\subsubsection{\label{sec_embedding}Time-series features}

\ourmethod{} tracks non-stationary dynamical variation in a time series by applying dimension reduction to a set of time-series features extracted from a sequence of time-localized windows of the original time series.
Our feature extraction procedure is as follows, starting with the case in which the time series is univariate.
We segment the observed time series $x_t^{(i)}$ of length $T$ samples into $N$ windows of length $W$ samples, obtaining an ordered sequence of windows $(\textbf{w}_1,\dots, \textbf{w}_{N})$.
In the case of non-overlapping contiguous windows, we have $N = \lfloor T/W \rfloor$, and for overlapping windows with stride $S$ we have $N = 1 + \lfloor (T - W)/S \rfloor$, where stride specifies the number of time series samples between the start of each window.
\ourmethod{} employs a set of $K$ functions $\{f_1,\dots,f_{K}\}$, where $f_j:\mathbb{R}^W \rightarrow \mathbb{R}$, i.e., time-series features, which are used to summarize dynamical variation in the time series of interest.
We apply each function to each window to obtain a feature matrix $F \in \mathbb{R}^{N \times K}$:
\begin{equation}
\label{eq_F}
F^{(i)} =
\begin{bmatrix}
    f_1(\boldsymbol{w}_1) & \dots& f_K(\boldsymbol{w}_1) \\
    f_1(\boldsymbol{w}_2) & \dots & f_K(\boldsymbol{w}_2) \\
    \vdots & \ddots & \vdots \\
    f_1(\boldsymbol{w}_N) & \dots & f_K(\boldsymbol{w}_N)
\end{bmatrix}\,.
\end{equation}
For a multivariate time series of dimension $D$ we concatenate the feature matrices $F^{(i)}$ to form $F = \begin{bmatrix}F^{(1)} & F^{(2)} & ... & F^{(D)} \end{bmatrix}$ so that $F \in \mathbb{R}^{N \times KD}$ [as depicted in Fig.~\ref{fig_summary}(d)].

The ability of \ourmethod{} to reconstruct a TVP depends on the set of $K$ time-series statistics being jointly informative of the statistical changes induced by the underlying TVP.
As noted in \citet{guttlerReconstructionParameterSpaces2001}, choosing an optimal time-series summary statistic requires knowledge of the process or TVP, and a similar logic applies to the set of candidate statistics to be used here.
For example, if training data with ground truth TVP(s) are available, then optimal features can be learned, e.g., from comparison across a comprehensive set of candidate time-series features (such as \textit{hctsa} \cite{fulcherHctsaComputationalFramework2017} or \textit{theft} \cite{hendersonFeatureBasedTimeSeriesAnalysis2025}).
Here, since we assume no knowledge of the underlying process, our approach is to use a diverse, representative set of time-series features that exhibit sensitivity to a broad range of dynamical properties that may potentially be affected by the underlying parameter(s).
We use the \textit{catch24} feature set, comprising a 22-feature subset of the \textit{hctsa} feature set \cite{lubbaCatch22CAnonicalTimeseries2019}, in addition to the mean and standard deviation.
Of note, the \textit{catch22} features are computed on z-scored data for each window, so the mean and standard variation add back in distributional information that is lost through z-scoring, which is important considering our prior work showing that the time-varying mean and variance constitute strong baselines on some \problemname{} problems \cite{owensParameterInferenceNonstationary2024}. 
We chose the \textit{catch24} feature set because it is concise, interpretable, computationally efficient, empirically derived, and contains a variety of statistics, including categories such as distributional shape (e.g., \texttt{mode\_10}, which returns the mode of a 10-bin histogram using z-scored data), linear autocorrelation structure (e.g., \texttt{low\_freq\_power}, which returns the relative power in the lowest 20\% of frequencies in the power spectrum), and nonlinear autocorrelation structure (e.g., \texttt{trev}, which returns the average of the cube of single-step temporal differences), among others.
However, the main results are not sensitive to the choice of feature set; we found broadly similar results using other time-series features sets, such as quantiles or time-variation in the power spectrum obtained using Fourier transformation (see Appendix~\ref{appendix_multiple}).

\subsubsection{\label{sec_dim_red}Dimension reduction via SFA}

How can we detect slow non-stationary variation by analyzing a matrix $F$ of time-varying statistical features, especially when some of these features may be related to a TVP while others are irrelevant?
Dimension-reduction methods are appropriate here, since our goal is to obtain a lower-dimensional transformation of $F$ that tracks one or more TVPs [Figs~\ref{fig_summary}(d)-(e)].
However, a problem that arises when designing a \problemname{} algorithm using dimension reduction and a large set of time-series features is that potentially only a small minority of features will be sensitive to a slow TVP.
In such cases, the TVP may be best approximated by a low-variance component that is difficult to detect using conventional dimension-reduction methods like PCA, which identifies directions of maximal shared variance within a dataset.
Moreover, pairs of time-series features may be correlated by construction, thereby biasing low-dimensional components in such directions, caused by the construction of the feature set rather than the structure of the   data\cite{fulcherInterpretableModelfreeInference2026}.
Indeed, \citet{harrisInferringParametricVariation2021} observed that such correlated features bias the output of dimension-reduction methods such as PCA, and that this bias can lead to the failure of TVP inference.
Our proposed solution to this problem is to instead use a dimension-reduction method that is constrained by temporal structure, with a bias towards slowness.


There are a number of dimension-reduction methods that have an \textit{inductive bias} (i.e. a constraint on the hypothesis space of possible patterns that can be extracted \cite{shalev-shwartzUnderstandingMachineLearning2014}) towards slowness (or related properties such as autocorrelation and local predictability), increasing their sensitivity to slow variation.
Elsewhere, we have referred to these as time-series dimension reduction (TSDR) methods, compared to general dimension-reduction methods such as PCA that are invariant to temporal permutation of the input data \cite{owensTimeseriesDimensionReduction}.
Of the various TSDR methods that aim to extract slowly varying components, here we select slow feature analysis (SFA) because it is a well-understood method for which existing implementations are available and which has previously demonstrated utility for \problemname{} when applied to time-delay-embedded data \cite{wiskottSlowFeatureAnalysis2002, wiskottEstimatingDrivingForces2003}.
Below, our selection of SFA is further supported through a numerical comparison with PCA (in Sec.~\ref{feature_feature}).


Slow feature analysis is a dimension-reduction method that extracts slow components from a multivariate time series, where slowness is operationalized by minimizing the square of the first derivative.
Starting with a time series $X$, SFA finds a linear transformation $Q$ that when applied to $X$ (as $XQ$) yields components ordered in terms of `slowness', based on an ordering of components by the least eigenvalues \cite{wiskottSlowFeatureAnalysis2002}.
Let $Z = XQ$ with variables $z_t^{(i)}$, then SFA constructs $Q$ so that $\langle(z_{t+1}^{(i)} - z_t^{(i)})^2 \rangle_t$ is minimized (that is, as slow as possible), under the constraints $\langle z_t^{(i)} \rangle_t = 0$ (zero mean), $\langle(z_t^{(i)})^2 \rangle_t = 1$ (unit variance), and $\langle z_t^{(i)}z_t^{(j)}\rangle_t = 0$, $\forall i \neq j$ (decorrelation), where $\langle \cdot \rangle_t$ is the time average of a variable \cite{wiskottSlowFeatureAnalysis2002}.
Without loss of generality, we assume that $X$ has been PCA whitened \cite{hyvarinenIndependentComponentAnalysis2001} so that it is decorrelated, with respect to the transformed variables with zero mean and unit variance.
The linear transformation $Q$ is then obtained by applying PCA (or singular value decomposition, SVD) to the time series of discrete first temporal derivatives of $X$ (computed using the first difference), but we select the least components/eigenvalues, representing directions in which these derivatives are minimized.
To perform \ourmethod{}, we apply SFA directly to our matrix $F$ of statistical time-series features [as depicted in Fig.~\ref{fig_summary}(d)].
Provided that the parameter variation is sufficiently slow and at least one statistic in our set of features is sensitive to parameter-driven dynamical variation, we expect the first component of SFA applied to $F$ to correlate with the TVP $\theta_t$ of the underlying non-stationary process [as depicted in Fig.~\ref{fig_summary}(e)].

\subsection{\label{sec_other_methods}Comparison methods}

To compare the performance of \ourmethod{}, we selected several other methods from the \problemname{} literature for comparison.
Of note, a persistent barrier to \problemname{} method comparison is the lack of open source implementations.
Accordingly, we were required to implement each comparator method, and thus chose methods that could be accurately implemented based on their description in the literature.
These working implementations are provided in the open code that accompanies the paper.
We also aimed to provide coverage of the different \problemname{} method categories, namely, statistical feature-based, dimension-reduction, prediction error, and phase-space partitioning methods \cite{owensParameterInferenceNonstationary2024}.

\begin{enumerate}

\item \textit{f-SFA using mean and variance (\baseline{}):}
This approach is an implementation of \ourmethod{} that uses only two features, mean and variance, to characterize the dynamics in each window (i.e., $K = 2$; rather than the main implementation analyzed here with $K = 24$).
\baseline{} is designed to serve as a simple and interpretable baseline method, leveraging our previous observation that mean and standard deviation successfully track parameter variation in the Lorenz and logistic map processes, respectively \cite{owensParameterInferenceNonstationary2024}.
This is a `statistical feature-based' \problemname{} method (in the schema of \cite{owensParameterInferenceNonstationary2024}).

\item \textit{Quadratic slow feature analysis (SFA2):}
Following \citet{wiskottEstimatingDrivingForces2003}, given a time series, we applied time-delay embedding with delay $\tau$ and dimension $m$, and quadratic polynomial basis expansion, followed by SFA to obtain a parameter estimate.
The time delay $\tau$ was set to the first zero-crossing of the autocorrelation function (and set to the minimum value across all variables of their respective first zero-crossing).
Using Wiskott's heuristic \cite{wiskottEstimatingDrivingForces2003}, $m$ was set by searching the integers $m = 1, \dots, 20$, selecting the value of $m$ for which the mean squared value of the first temporal derivative of the first SFA component was minimized.
This is a `dimension-reduction' \problemname{} method \cite{owensParameterInferenceNonstationary2024}.

\item \textit{Echo state network prediction error (ESN):}
Inspired by the approach of \citet{gunturkunSequentialReconstructionDrivingforces2010}, given a time series, 50 echo state networks (ESN) of internal dimension 30 were trained to minimize the one-step prediction error and then a time series of the mean prediction error was obtained by averaging across the 50 networks.
Rather than adaptive filtering, we smoothed the resulting signal with a Gaussian convolutional filter to obtain a TVP estimate.
We used $\sigma = 166$ for the Gaussian kernel standard deviation parameter, yielding a filter window of approximately $10^3$ steps between $\pm 3\sigma$ to maintain consistency with the window sizes used in the \ourmethod{} and CD methods.
This is a `prediction-error' \problemname{} method \cite{owensParameterInferenceNonstationary2024}.

\item \textit{Characteristic distance (CD):}
Following the general approach of \citet{nguyenNewInvariantMeasures2015}, given a time series, 100 random points were uniformly sampled from a phase-space hypercube bounding the time-series points, but with boundaries extended by $\pm 10\%$ in each dimension to allow for sampling of points `outside' the distribution.
For each time-series point, a vector of Euclidean distances was computed from each of the random points, resulting in a time series $M \in \mathbb{R}^{T \times 100}$.
Mean distances were then calculated using sliding windows of size $10^3$ along the columns of $M$, tracking mean deviation from the sampled points over time.
Finally, SFA was applied to obtain a TVP estimate.
This is a `phase-space partitioning' \problemname{} method \cite{owensParameterInferenceNonstationary2024}.

\end{enumerate}

\subsection{\label{sec_nonstat_procs}Non-stationary processes}

We selected several chaotic flows and maps to serve as non-stationary processes to compare the performance of \ourmethod{} and the other \problemname{} methods.
We chose to include the Lorenz \cite{lorenzDeterministicNonperiodicFlow1963} and logistic map \cite{maySimpleMathematicalModels1976} processes because these are two of the most commonly studied comparison problems in the \problemname{} literature.
However, we previously showed that \problemname{} problems involving these systems are `trivially' solvable using simple baseline algorithms, e.g., using local means to track the $\rho$ parameter of the Lorenz process and local standard deviations to track the $r$ parameter of the logistic map \cite{owensParameterInferenceNonstationary2024}.
To discover more challenging problems, following the approach detailed in \citet{owensParameterInferenceNonstationary2024}, we manually searched through the chaotic processes collated by Sprott \cite{sprottChaosTimeSeriesAnalysis2001} (comprising 29 discrete maps and 33 continuous flows) and Gilpin \cite{gilpinChaosInterpretableBenchmark} (comprising 131 flows).
Specifically, we searched for systems for which TVP inference was possible using at least one of our considered \problemname{} methods (Sec.~\ref{sec_other_methods}), but where inference failed for \baseline{}.
We found four systems that satisfied these characteristics: two chaotic flows, the Blasius process \cite{blasiusComplexDynamicsPhase1999} and the Langford process \cite{langfordNumericalStudiesTorus1984}; and two chaotic maps, the sine map \cite{sprottChaosTimeSeriesAnalysis2001} and the predator-prey map \cite{beddingtonDynamicComplexityPredatorprey1975}.

To compare the performance of \ourmethod{} against the other methods, we simulated each system under the influence of one (or more) TVP(s), yielding non-stationary time series.
For each system, the TVPs served as sources of non-stationarity with known variation that we compared each method against.
Time-series realizations of each process were generated from the model equations, either through iteration of a map
$\textbf{x}_{t+1} = G(\textbf{x}_t, \theta_t)$
or integration of a flow
$\frac{d}{dt}\textbf{x} = g(\textbf{x}, \theta(t))$, given some initial state $\textbf{x}_0$ and some time-varying parameter $\theta(t)$ or $\theta_t$ (where $G$ and $g$ for each system is defined in Appendix~\ref{appendix}).
Given a total integration time (or iteration number) $T_{\text{max}}$, we investigated two different functional forms for the time-variation of parameters: a single-period sinusoid,
\begin{equation}
\label{eq_sinusoid}
    \phi(t) = \sin(2\pi t/T_{\text{max}})\,,
\end{equation}
and Wiskott's sum of three sinusoids
\cite{wiskottEstimatingDrivingForces2003},
\begin{equation}
    \label{eq_wiskott}
    \begin{gathered}
        \text{$\phi(t) = \sin(2\pi (5t/T_{\text{max}})) + \sin(2\pi (11t/T_{\text{max}}))$} \\
        \text{$+ \sin(2\pi(13t/T_{\text{max}}))\,.$}
    \end{gathered}
\end{equation}
In both cases, the resulting TVP was:
\begin{equation}
    \label{eq_tvp}
    \theta(t) = \theta_0[1 + \alpha \phi(t)]\,,
\end{equation}
where $\theta_0$ is the default parameter value, and $\alpha$ controls the amplitude of variation about the default value.
Unless otherwise specified, we set $\alpha = 0.1$, corresponding to a variation of $\pm 10\%$ from the value of the default parameter value $\theta_0$.
The various process models and default parameter values that were used are detailed in Appendix~\ref{appendix}.

For integration, we used the Runge--Kutta--Fehlberg (RKF45) method \cite{fehlbergLoworderClassicalRungeKutta1969} using the SciPy library \cite{virtanenSciPyFundamentalAlgorithms2020}.
The integration of chaotic flows was performed over $10^3$ time units, which we will refer to as seconds, with a uniform sampling period $\Delta t = 0.01\,\textnormal{s}$ (unless otherwise specified).

To simulate observation noise, Gaussian noise was added to each individual time-series channel.
We calculated the signal-to-noise ratio (SNR) per channel as:
\begin{equation}
    \label{eq_snr}
10 \log_{10}(\sigma^2_s/\sigma^2_n)\,,
\end{equation}
where $\sigma^2_s$ is the variance of the signal and $\sigma^2_n$ is the variance of the noise.
The performance of each method was quantified by comparing the inferred TVP $\hat \theta_t$ with the ground-truth $\theta_t$ using Pearson $R^2$ \cite{rodgersThirteenWaysLook}.
Unless otherwise specified, we used a window length of $W = 1000$ samples and a stride length of $S = 200$ samples for all methods, striking a balance between temporal resolution and computation time.
Our investigations include evaluating each method's performance across many of the hyperparameters above, including robustness to noise noise (SNR), sampling interval ($\Delta t$), samples per window ($W$), and the amplitude of parameter variation ($\alpha$).

\section{\label{sec_results}Results}

In what follows, we detail the results of a series of numerical experiments that evaluate the performance of \ourmethod{}, compare it to other \problemname{} methods, and investigate how it is affected by various parameters.
First, in Sec.~\ref{feature_feature}, we show that \ourmethod{} is robust to the presence of correlated statistical features, which is not the case when PCA is substituted for SFA.
Then, in Sec.~\ref{experiment_comparative}, we show that the performance of \ourmethod{} is superior to our set of comparator algorithms, covering several \problemname{} categories, when compared across a range of non-stationary systems and levels of observation noise.
Finally, we present numerical experiments showing how the performance of \ourmethod{} varies in more challenging settings in relation to factors such as parameter timescale (Sec.~\ref{experiment_timescale}), parameter amplitude (Sec.~\ref{experiment_amplitude}), and generalization to unseen parameter values (Sec.~\ref{experiment_ts_length}).
The code to reproduce each experiment is available on the GitHub repository: \url{https://github.com/KieranOwens/fsfa_paper}.

\begin{figure*}
\includegraphics[width=1.0\textwidth]{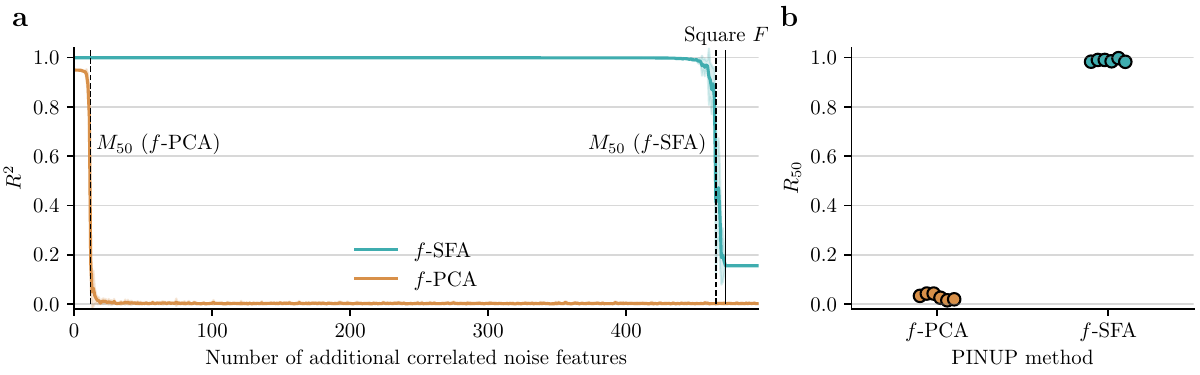}
\caption{\label{fig:feat-feat}
\textbf{\ourmethod{} performs well in the presence of an increasing number of correlated noise features whereas the performance of $f$-PCA degrades.}
Correlated noise features were added incrementally to observe the effect on PINUP performance for both \ourmethod{} and $f$-PCA, with performance measured via Pearson correlation ($R^2$) with the ground truth TVP.
\textbf{a.} A case study of \ourmethod{} versus $f$-PCA performance for the logistic map shows that $f$-PCA performance degrades with only a small number of correlated features, whereas \ourmethod{} performs well until algorithmic failure occurs when the number of features approaches the number of samples in the feature matrix;
\textbf{b.} A comparison of $R_{50} = M_{50}/(N - DK)$, where $M_{50}$ is the number of redundant features at which $R^2$ performance falls by $50\%$ and where $(N-DK)$ corrects for the size of the feature matrix $F$ (see Eq.~\eqref{eq_F}).
}
\end{figure*}

\subsection{\label{feature_feature}Feature--feature correlation}

Two key principles underlie our design of the \ourmethod{} algorithm.
Firstly, a large time-series feature set is used with the goal of increasing sensitivity to a wide range of potential sources of statistical variation through which non-stationarity may manifest.
Secondly, SFA is used for its strong inductive bias towards slowness, so that any given TVP is not drowned out by a large number of irrelevant correlated features, as can occur with variance-based dimension reduction.
To verify the robustness of \ourmethod{} to the presence of correlated features relative to $f$-PCA (i.e., \ourmethod{} but with SFA replaced by PCA) we devised the following numerical experiment.
For each of our six chaotic non-stationary systems (the logistic, sine, and predator-prey maps, and the Lorenz, Blasius, and Langford processes), driven by a single-period sinusoidal parameter, \textit{catch24} statistical features were computed across sliding windows, yielding feature matrices $F \in \mathbb{R}^{N \times DK}$ (see Eq.~\eqref{eq_F}).
Each $F$ was then $z$-score normalized and augmented with increasing numbers of correlated noise features prior to application of SFA and PCA, respectively.
A baseline noise vector $\boldsymbol{\mu}^{(\text{master})} \in \mathbb{R}^N$ was sampled i.i.d. as  $\mu^{(\text{master})}_n  \sim \mathcal{N}(0,1)$.
By perturbing $\boldsymbol{\mu}^{(\text{master})}$ with small amounts of noise, correlated vectors $\boldsymbol{\mu}^{(i)}$ were then generated so that $\mu^{(i)}_n = \mu^{(\text{master})}_n + \xi^{(i)}_n$, where $\xi^{(i)}_n \sim \mathcal{N}(0,0.1)$, yielding an augmented feature matrix:

\begin{equation}
    \label{eq_F_aug}
    \tilde{F}_m =
    \begin{bmatrix}
        F & \boldsymbol{\mu}^{(1)} & \dots & \boldsymbol{\mu}^{(m)}
    \end{bmatrix} \,.
\end{equation}
For each system, for each number of correlated noise vectors $m$, \ourmethod{} and $f$-PCA performance was quantified using $R^2$ correlation with the ground truth TVP, averaged over $10$ iterations of the above process.

Our results for the logistic map, shown in Fig.~\ref{fig:feat-feat}(a), provide an illustrative case study of the different behavior of the \ourmethod{} and $f$-PCA algorithms in the presence of correlated features.
Good performance ($R^2 > 0.9$) was seen when $f$-PCA was applied to the unaugmented feature matrix $F$, but its performance deteriorates rapidly as the number of correlated noise features approaches $m = 15$.
This supports the claim that variance-based dimension reduction using $f$-PCA is indeed affected by the composition of the feature set, with performance that degrades in the presence of high-magnitude shared variance that is not related to the underlying TVP.
In contrast, the performance of \ourmethod{} remained excellent ($R^2 > 0.99$) up to much higher values of $m\approx 450$, followed by a rapid drop in performance.
The drop in \ourmethod{} performance was seen as $m$ approached $N - DK$, i.e., as the number of features in $\tilde{F}_m$ approached the number of samples, resulting in rank deficient feature matrices and the failure of dimension reduction using SFA.
Thus, SFA successfully identifies the relevant features that track the TVP despite a `needle in a haystack' scenario involving a large number of irrelevant, correlated features.
The inductive bias of dimension-reduction methods like SFA indeed facilitates the use of large, diverse feature sets for this problem, overcoming the limitations of manually selecting statistical features, as discussed by \citet{guttlerReconstructionParameterSpaces2001}.

To compare the performance of \ourmethod{} versus $f$-PCA across the six non-stationary processes we tested, we used $M_{50}$, defined as the minimum number of correlated noise features at which performance was degraded by at least $50\%$.
Considering that dimensionality differed between our selected chaotic systems (e.g., the logistic map is 1-dimensional and the Lorenz process is 3-dimensional), we normalized $M_{50}$ to construct the performance metric $R_{50} = M_{50} /(N - DK)$, accounting for the number of noise features required to produce a square feature matrix.
The values of $R_{50}$ for each system are shown in Fig.~\ref{fig:feat-feat}(b).
As was observed for the logistic map case study, the performance of \ourmethod{} deteriorates for each system as the number of correlated features $m$ approaches $N - DK$, whereas $f$-PCA fails much earlier.

Collectively, these experiments support our claim that $f$-PCA is susceptible to failure in the presence of spurious feature--feature correlation (consistent with the findings of \citet{harrisInferringParametricVariation2021}), whereas \ourmethod{} is robust to such effects when the variation in the underlying parameter is slow, matching the inductive bias of SFA.
This overcomes a major barrier to the use of larger time-series feature sets, which we expect to be able to track a wider range of non-stationary dynamical variation, and supports our choice of SFA for dimension reduction.

\begin{figure*}
\includegraphics[width=1.0\textwidth]{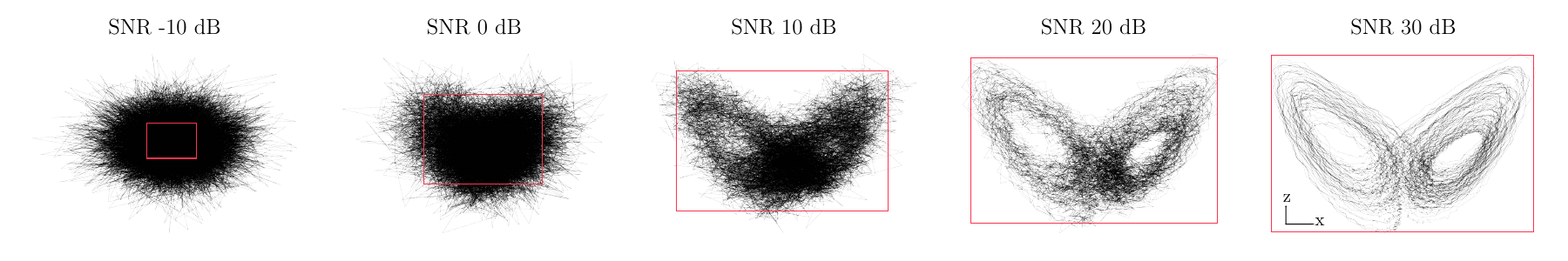}
\caption{\label{fig:noise}
\textbf{Observation noise:} examples of different levels of observation noise applied to the Lorenz process are shown.
The red rectangle is used to denote the same region of phase space in each case, demonstrating the approximate scale over which the deterministic dynamics evolve (and the extent to which observation increasingly dominates at lower SNR).
}
\end{figure*}


\subsection{\label{experiment_comparative}Comparative performance}

\begin{figure*}
\begin{center}
\begin{minipage}[t]{0.48\textwidth}
\includegraphics[width=\linewidth]{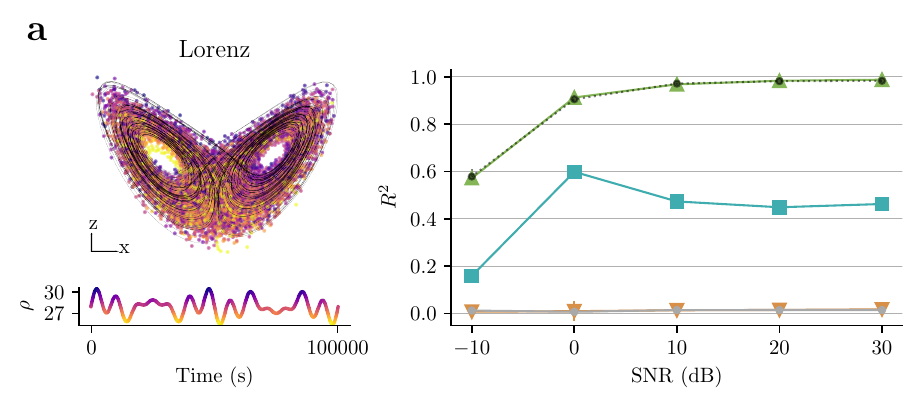}
\end{minipage}\hfill
\begin{minipage}[t]{0.48\textwidth}
\includegraphics[width=\linewidth]{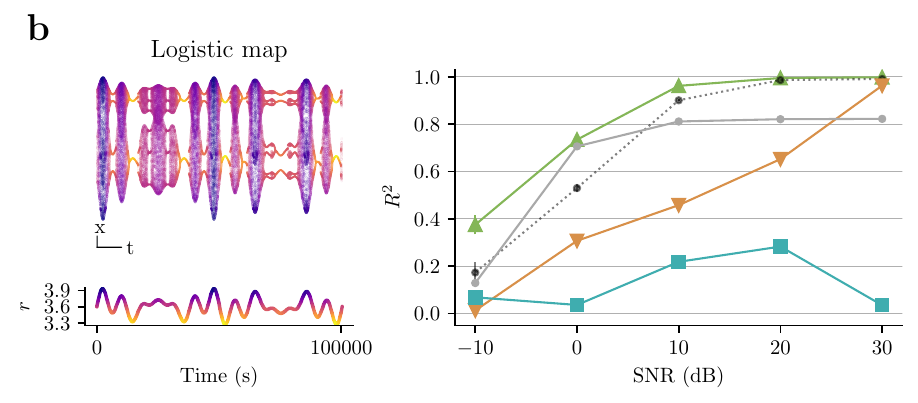}
\end{minipage}
\end{center}
\begin{center}
\begin{minipage}[t]{0.48\textwidth}
\includegraphics[width=\linewidth]{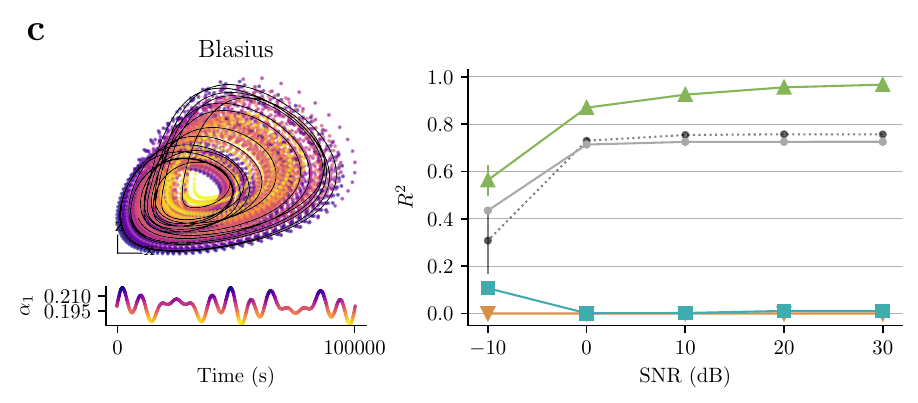}
\end{minipage}\hfill
\begin{minipage}[t]{0.48\textwidth}
\includegraphics[width=\linewidth]{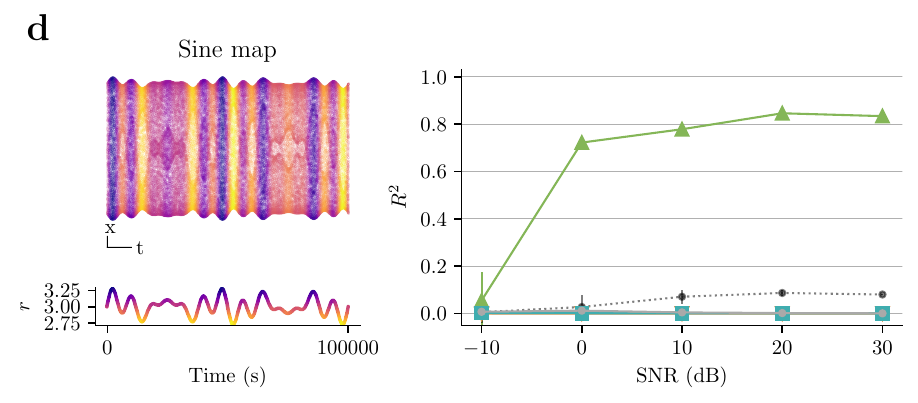}
\end{minipage}
\end{center}
\begin{center}
\begin{minipage}[t]{0.48\textwidth}
\includegraphics[width=\linewidth]{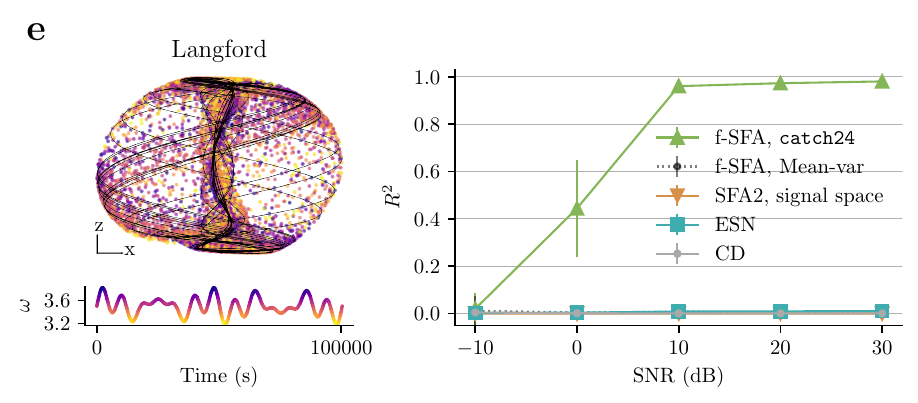}
\end{minipage}\hfill
\begin{minipage}[t]{0.48\textwidth}
\includegraphics[width=\linewidth]{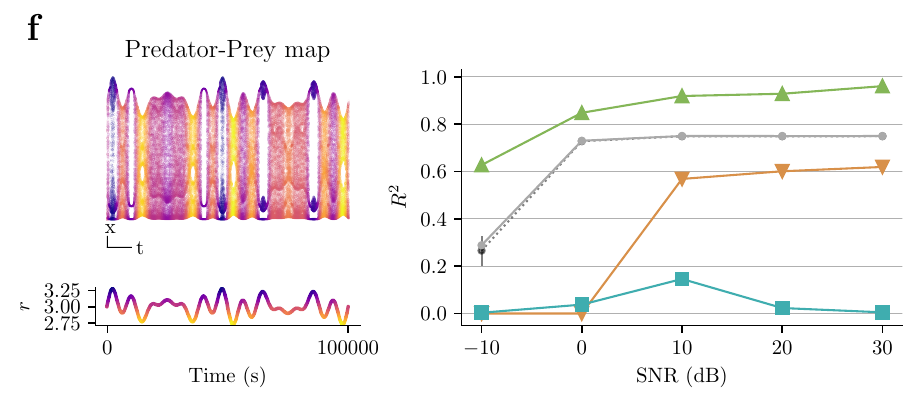}
\end{minipage}
\end{center}
\caption{\label{fig:comparative}
\textbf{\ourmethod{} exhibits strong and more noise-robust performance compared to other \problemname{} methods.}
Each panel displays results from a parameter inference experiment, where the process is depicted on the left with the time-varying parameter below.
Note that the chaotic flows [(a), (c), (e)] are visualized as 2-dimensional projections whereas the maps [(b), (d), (f)] are visualized as the time series of a single labeled variable.
For each experiment, we plot the signal-to-noise ratio (SNR) against performance (mean Pearson $R^2$, with error bars indicating standard deviation) for each \problemname{} method: \ourmethod{}, feature-based slow feature analysis (one version using \textit{catch24} features and the other using mean and variance); SFA2, quadratic slow feature analysis; ESN, echo state network prediction error; and CD, characteristic distance (cf. Sec.~\ref{sec_other_methods} for details).
The processes (and varied parameters) are
(a) the Lorenz process (parameter $\rho$),
(b) the logistic map (parameter $r$),
(c) the Blasius process (parameter $\alpha_1$),
(d) the sine map (parameter $r$),
(e) the Langford process (parameter $d$), and
(f) the predator--prey map (parameter $r$).
The legend for panels (a)-(f) is located in panel (e).
}
\end{figure*}

We next performed numerical experiments comparing the \problemname{} performance of: (i) \ourmethod{} (using $\textit{catch24}$ features), (ii) \baseline{}, (iii) SFA2, (iv) ESN, and (v) CD (see Sec.~\ref{sec_other_methods}).
We wanted to determine whether, despite its simplicity, the performance of \ourmethod{} would match or exceed that of the other \problemname{} methods.
These methods were each applied to time series of length $T = 10^5$ samples simulated from each of three chaotic maps (Eq.~\eqref{eq_logistic}-\eqref{eq_pred_prey}) and three chaotic flows (Eq.~\eqref{eq_lorenz}-\eqref{eq_Blasius}) according to the methodology outlined in Sec.~\ref{sec_nonstat_procs}, using Wiskott's sum of sinusoids (Eq.~\eqref{eq_wiskott}) for the TVP [e.g., see the time course of the $\rho$ parameter in Fig.~\ref{fig:comparative}(a)].
Performance was then averaged over multiple iterations of different levels of measurement noise, indexed by SNR, from -10\,dB through 30\,dB.
The range of SNR that we investigated is shown in Fig.~\ref{fig:noise} for a non-stationary Lorenz process.
At a SNR of 30\,dB, the attractor and dynamics are well-preserved, whereas at SNR -10\,dB the underlying attractor is visually completely obscured by noise, presenting a highly challenging problem.


The results are shown in Fig.~\ref{fig:comparative}, which for each system shows the process, the TVP, and a plot of performance versus SNR for each of the five \problemname{} methods we analyzed.
\ourmethod{} achieved equivalent or superior performance to the other methods across all systems and noise levels, and achieved $R^2 > 0.8$ at the lowest noise level for all systems.

The performance of all methods deteriorated with increasing noise, but \ourmethod{} was the most robust to noise.
Specifically, \ourmethod{} achieved high-quality TVP reconstructions ($R^2 > 0.9$) for 5 of our 6 systems across the SNR range $10$ to $30$\,dB, and even achieved $R^2 > 0.8$ at SNR 0\,dB for the Lorenz, Blasius, and Predator-Prey systems (Fig.~\ref{fig:comparative}).
We conjecture that the noise-robustness of \ourmethod{} is due to the robustness of the underlying statistics, many of which are capable of producing reliable estimates in the presence noise (such as mean, variance, and the centroid of the power spectrum), given sufficiently long windows.

For the Lorenz and logistic map systems, strong performance was achieved by our baseline method, \baseline{}, which uses just two features, the mean and variance (Sec.~\ref{sec_other_methods}).
For such systems, if the sampling rate is high enough and the variation is slow enough that good local statistical estimates can be computed in each window, simple sliding-window estimates of mean and variance can track the TVP, as we previously showed for the `easy' problems of tracking Lorenz parameter $\rho$ and the logistic map parameter $r$ \cite{owensParameterInferenceNonstationary2024}.
Our results here are consistent with this finding, with \baseline{} performing very well for the Lorenz and logistic map processes at high SNRs [Figs~\ref{fig:comparative}(a)-(b)].
Variation in the $\rho$ parameter causes translation of the Lorenz attractor along the $z$-axis, which is why the sliding-window mean can be successfully used for TVP inference.
For the logistic map, the range of time-series values increases monotonically with the parameter $r$ over the range of TVP values in this experiment, so the TVP could be tracked using sliding-window variance alone.

In contrast to the Lorenz and logistic map processes, other systems were more challenging.
For the sine map [Fig.~\ref{fig:comparative}(d)], \ourmethod{} achieved $R^2>0.8$ for SNRs $\geq 20$\,dB, and on the Langford process [Fig.~\ref{fig:comparative}(e)], \ourmethod{} achieved $R^2>0.9$ for SNRs $\geq 10$\,dB, while all other \problemname{} methods failed on these two systems.
The fact that \baseline{} failed in these cases suggests that part of why these problems pose a greater challenge for \problemname{} is that they cannot be solved by tracking the first two moments of the distribution.
In contrast, the success of \ourmethod{} shows that statistical properties may still exist that vary approximately monotonically with the TVP, allowing accurate TVP reconstruction (again, given a sufficiently high sampling rate and sufficiently slow TVP variation).
Our results suggest that the use of a diverse set of time-series features (even the relatively compact set, \textit{catch24}, used here) improves the sensitivity of \ourmethod{} to different kinds of parameter-driven dynamical variation.
To further examine the sensitivity of \ourmethod{} to the choice of features, we repeated our experiment using other feature sets in place of \textit{catch24}, such as quantiles and spectral power of frequencies computed using the short-time Fourier transform (STFT).
The compact \textit{catch24} feature set performed well overall, however the optimal feature set depended on the particular problem, and in some cases an improvement in performance could be obtained by combining the feature sets (see Appendix~\ref{appendix_results}).

Empirically, the performance of \baseline{} was almost always superior to that of SFA2, ESN, and CD.
The exception was the marginal superiority of CD in certain high-noise settings (e.g., the logistic map at SNR 0\,dB and the Blasius process at SNR -10\,dB), noting that CD utilizes mean distances from multiple random points in phase space, yielding information similar to that of the sliding-window mean.
The superiority of \baseline{} suggests that SFA2, ESN, and CD may depend on distributional properties for their performance on \problemname{} problems and strengthens the case for using \baseline{} as a simple baseline comparison to provide a basic demonstration of a performance improvement when developing new algorithms.
Additionally, given that the performance of \ourmethod{} was superior or equivalent to \baseline{} under all experimental conditions, this suggests that \ourmethod{} represents a methodological advance for PINUP problems involving slow TVPs.

The above experiment only considered the case of a single TVP for each process, but in principle \ourmethod{} can be used to infer multiple TVPs.
While the single-TVP case is the focus of this paper, we provide a simple case study demonstrating the ability of \ourmethod{} to simultaneously inferring three TVPs for the Lorenz process in Appendix~\ref{appendix_multiple}.
While a fuller investigation of the multiple TVP setting is out of the scope of this work, we note the importance of future research to better test the feasibiliy of multi-parameter \problemname{} across a range of systems, and to investigate how the accuracy of inference depends on the differential timescales of the independent TVPs.

%


In summary, we found that \ourmethod{} achieves superior or equivalent \problemname{} performance and greater noise robustness, compared to \baseline{}, SFA2, ESN, and CD.
Difficult \problemname{} problems were identified for which TVP inference was only successful using \ourmethod{}.
In cases where multiple methods performed well, \baseline{} also exhibited strong performance, consistent with the suitability of this method as a baseline comparator when comparing novel \problemname{} algorithms.
The superior performance of \ourmethod{} over the comparator methods, in spite of its relative algorithmic simplicity, demonstrates the power of combining a broad set of time-series features with an inductive bias towards slowness via SFA.
To refine our understanding of the conditions under which \ourmethod{} performs best, we next turn our attention to the impact of parameter timescale.

\begin{figure*}
\begin{center}
\includegraphics[width=1.0\textwidth]{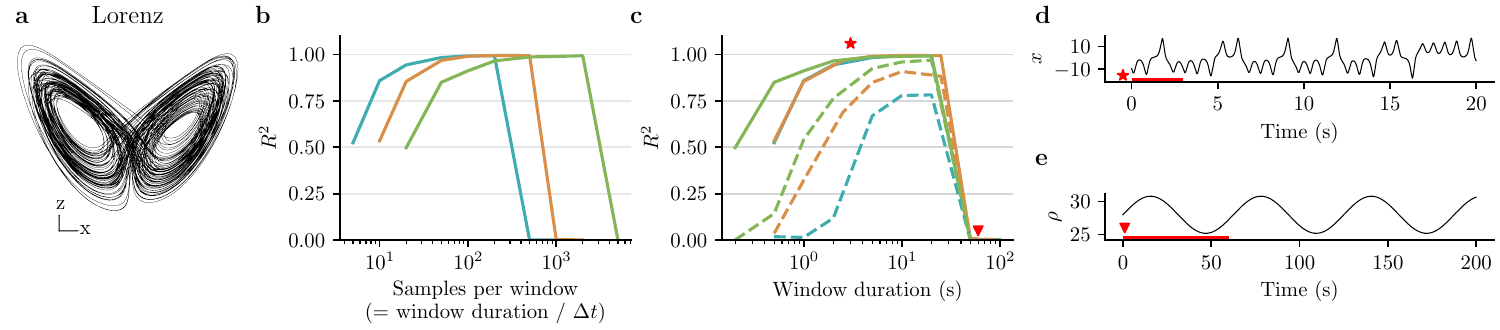}
\end{center}
\begin{center}
\includegraphics[width=1.0\textwidth]{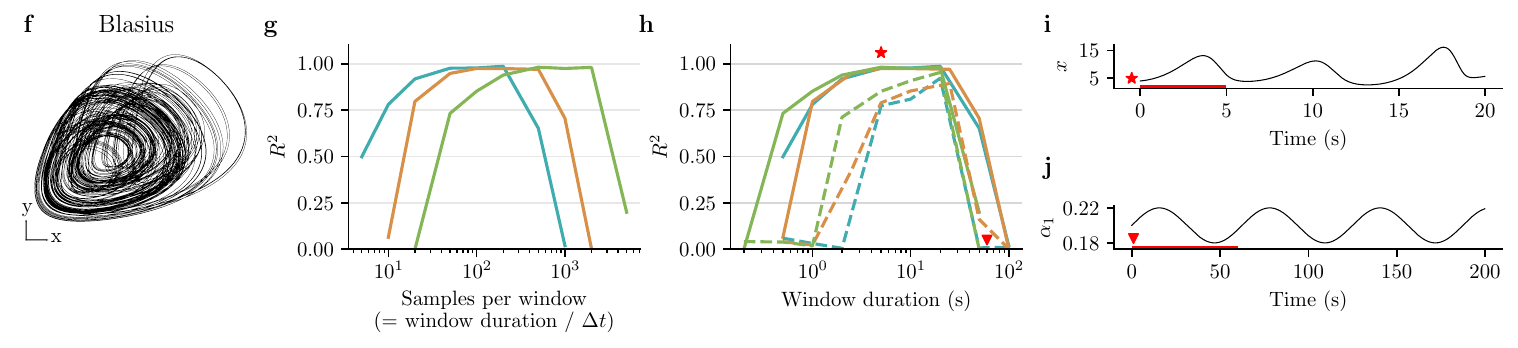}
\end{center}
\begin{center}
\includegraphics[width=1.0\textwidth]{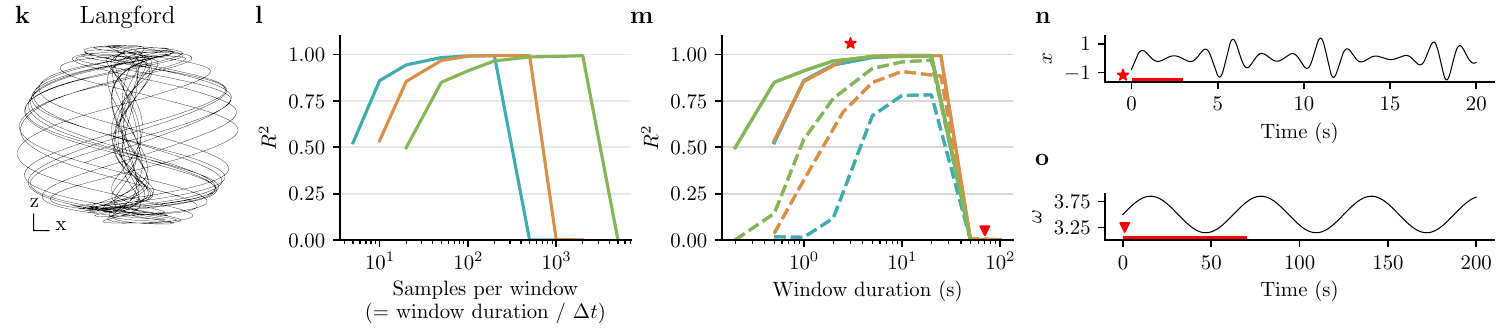}
\end{center}
\begin{center}
\includegraphics[width=1.0\textwidth]{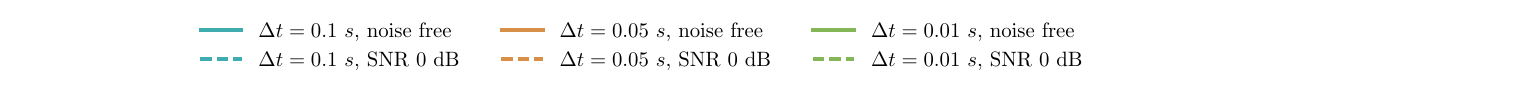}
\end{center}
\caption{\label{fig:timescale}
\textbf{The performance of \ourmethod{} depends on window duration relative to the timescale of the underlying process and the time course of the TVP driving non-stationarity.}
Non-stationary Lorenz, Langford, and Blasius processes [(a), (f), (k)] were simulated using a sinusoidal TVP.
The performance of \ourmethod{} is shown for samples per window [(b), (g), (l)] and window duration [(c), (h), (m)], with noise-free experiments are shown in solid lines and 0\,dB noise with dashed lines, and with sampling rate denoted by color.
In the noise-free setting, \ourmethod{} performance ($R^2$) was dependent on window duration rather than samples per window [(b)-(c), (g)-(h), (l)-(m)].
In the high-noise setting (SNR of 0\,dB), \ourmethod{} performance was better when the sampling rate was high ($\Delta t = 0.01\,s$), so that there were more samples in any window of a given duration [(c), (h), (m)).
The range of window durations over which \ourmethod{} was optimal was bounded below by a duration approximately corresponding to the duration of an orbit of the underlying process, and bounded above by the period length of the sinusoidal TVP [(c)-(e), (h)-(j), (m)-(o)].
The lower bounds and corresponding durations are shown with a red star and bar and the upper bounds and durations are shown with a red triangle and bars.
}
\end{figure*}



\subsection{\label{experiment_timescale}Parameter timescale}

Next, we aimed to understand the conditions underlying strong \ourmethod{} performance, starting with the parameter timescale.
Specifically, we must consider the timescales of the underlying process, the window over which a statistic is calculated, and the time course of the TVP that drives non-stationarity.
The critical hyperparameter with respect to the timescale is the sliding window length, which can be defined in terms of the number of time-series samples $W$ within the window, or the window duration $\tau_w = W \Delta t$, measured in integration seconds for continuous-time processes.
Ideally, we want a window that is long enough to obtain a good statistical estimate, but short enough that the dynamics remain approximately stationary over the duration of the window.
As the window duration increases, we approach a TVP timescale $\tau_p$ over which the TVP-driven dynamics of the process becomes non-stationary, leading to a failure of \ourmethod{}.
As the window duration decreases, it may fall below some threshold $\tau_s$ below which the statistic(s) modulated by the TVP cannot be reliably estimated.

For example, we may need to observe at least one or more orbits around a chaotic attractor to accurately estimate a property such as the mean position of the attractor in phase space or the fundamental frequency of the power spectrum; whereas a phase-space trajectory comprising only a small fraction of such an orbit would not be sufficient to accurately estimate the chosen statistic.
Furthermore, most statistics will be more sensitive to noise and therefore less robust when the number of samples $W$ is low.
Consequently, we hypothesized that in \problemname{} methods such as \ourmethod{} that involve windowing, the choice of $W$ involves a trade-off: if $W$ is too large, the windows will no longer be pseudo-stationary, while if $W$ is too small, there will not be sufficient samples to accurately estimate the statistics of the process.

To assess the hypothesized timescale-dependence of \ourmethod{} performance, we simulated non-stationary Lorenz, Langford, and Blasius processes under the influence of slow sinusoidal TVPs (Eq.~\eqref{eq_sinusoid}).
We considered continuous flows rather than discrete maps for this experiment because the duration $\Delta t$, and therefore $\tau_w$, can only be varied in the former case, imposing a timescale on statistical estimation.
Sinusoidal TVPs were chosen for this experiment because the sinusoid period provides a natural TVP timescale.
We examined the performance, assessed as $R^2$, of \ourmethod{} over a range of window lengths $W$, for sampling intervals of $\Delta t = 0.1$, $0.05$ and $0.1$\,s, and for both noise-free and SNR $0$\,dB observation noise conditions.
Performance was evaluated for the Lorenz [Fig.~\ref{fig:timescale}(a)], Blasius [Fig.~\ref{fig:timescale}(f)], and Langford [Fig.~\ref{fig:timescale}(k)] processes as a function of both the number of time-series samples per window $W$ [Figs~\ref{fig:timescale}(b), (g), and (l)] and the window duration $\tau_\text{window}$ [Figs~\ref{fig:timescale}(c), (h), and (m)].

The performance of \ourmethod{} was primarily dependent on the window duration $\tau_w$ rather than the number of samples per window $W$.
In noise-free conditions, across all three non-stationary processes, the optimal performance of \ourmethod{} was observed over the same range of $\tau_{\text{window}}$ for each sampling rate [Figs~\ref{fig:timescale}(c), (h), and (m)], while the optimal range of $W$ differed for each sampling rate [Figs~\ref{fig:timescale}(b), (g), and (l)].
This finding supports our hypothesis that $\tau_w$ is a critical hyperparameter for \ourmethod{} performance for processes sampled in continuous time.
In contrast, under the high-noise condition (SNR = 1), the performance was better when the sampling rate was high ($\Delta t = 0.01$\,s), consistent with our expectation that a higher number of samples per window ($W$) confers greater noise robustness when estimating time-series statistics.
In short, the choice of $\tau_w$ is a key determinant of \ourmethod{} performance and, for a fixed $\tau_w$, noise robustness improves at higher sampling rates.

For each process, there was a range of $\tau_w$ over which strong performance was obtained, falling between the points marked with red star ($\tau_w^\text{min}$) and red triangle ($\tau_w^\text{max}$) annotations in Figs~\ref{fig:timescale}(c), (h), and (m).
For each process, $\tau_w^\text{min}$ and $\tau_w^\text{max}$ are similarly annotated for the observed dynamics [Figs~\ref{fig:timescale}(d), (i), and (n)] and the TVP [Figs~\ref{fig:timescale}(e), (j), and (o)], respectively, with red bars indicating the durations.
For each process, observe that $\tau_w^{\text{min}}$ roughly corresponds to the duration of at least one orbit of the process, which is the least timescale over which the attractor is minimally sampled [Figs~\ref{fig:timescale}(d), (i), and (n)].
On shorter timescales, certain dynamical properties can no longer be accurately estimated, such as the mean position of the attractor in phase space, or the dominant frequency of the process.
This is the lower threshold $\tau_s$ at which statistical estimation breaks down.
Furthermore, observe that this threshold increases in the presence of noise [Figs~\ref{fig:timescale}(c), (h), and (m)], since more samples $W$ are required to obtain accurate statistical estimates from noisy data.

A deterioration in performance was also observed as $\tau_w$ increased beyond a certain threshold.
Specifically, for each process, $R^2 \rightarrow 0$ as the window duration approached $\tau_w^{\text{max}}$, which corresponds to the period of each sinusoidal TVP [Figs~\ref{fig:timescale}(e), (j), and (o)].
This is expected behavior because the TVP variation cannot be resolved from a window duration equal to the period of the TVP (and therefore `smears over' the relevant variation).
The knee of each graph [Figs~\ref{fig:timescale}(c), (h), and (m)], at which performance rapidly declines with increasing $\tau_w$, corresponds to the upper threshold $\tau_p$ at which TVP inference fails due to large variations in the TVP over the duration of a window.

In summary, our experiments indicate that the success of \ourmethod{} requires the duration $\tau_w$ to be sufficiently short that the process is approximately stationary (hence less than $\tau_p$) and long enough to contain a trajectory from which dynamical properties can be accurately estimated (hence greater than $\tau_s$ and with sufficiently many samples $W$ to withstand any effects of noise).
In practice, domain knowledge may be needed to select an optimal value of $\tau_w$, matching the timescale of interest to the scientific question being investigated.
Future work could investigate how to automatically optimize this hyperparameter, in order to address scientific questions about non-stationarity that manifests on a given timescale of interest.

\begin{figure*}
\includegraphics[width=500pt]{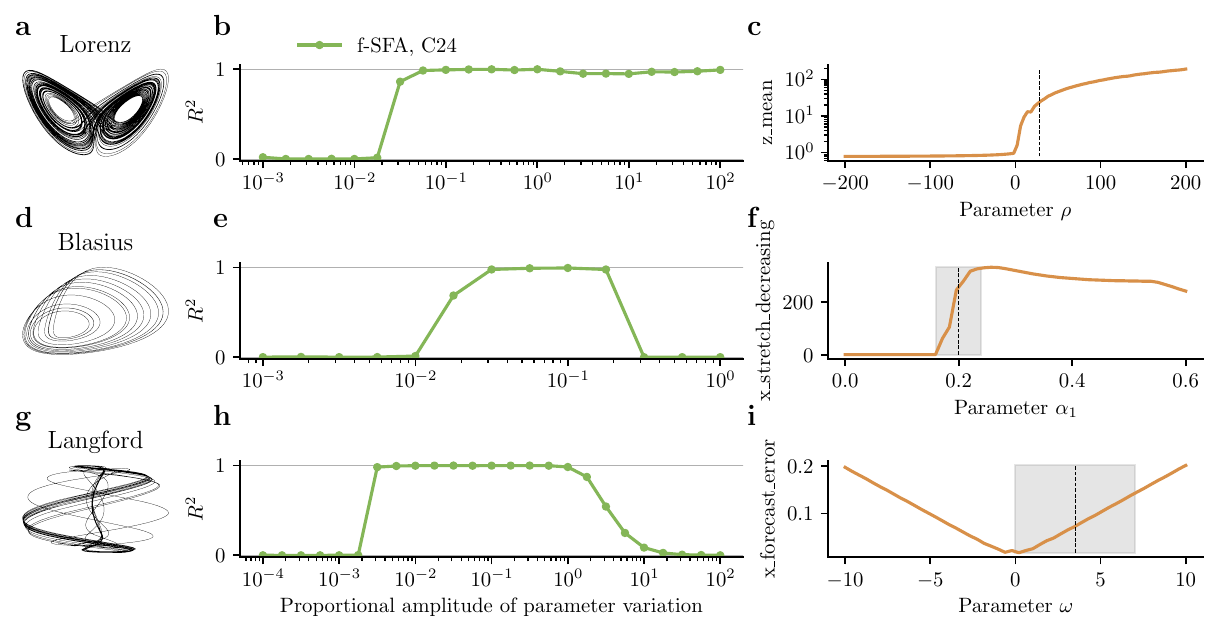}
\caption{\label{fig_amplitude}
\textbf{Good performance of \ourmethod{} requires a time-series statistic that is monotonically related to the TVP.}
The performance (Pearson $R^2$) of \ourmethod{} is shown for different proportional amplitudes of parameter variation (corresponding to values of $\alpha$ in Eq.~\eqref{eq_tvp}) using noiseless observations of the Lorenz [(a)-(b)], Blasius [(d)-(e)], and Langford [(g)-(h)] systems driven by slow sinusoidal TVPs.
The empirical relationship between TVP values and an illustrative time-series statistic is shown for each process, including the Lorenz parameter $\rho$ and the mean of the $z$ variable (c), the Blasius parameter $\alpha_1$ and \textit{catch22} feature \texttt{stretch\_decreasing} applied to the $x$ variable (f), and the Langford parameter $\omega$ and \textit{catch22} feature \texttt{forecast\_error} applied to the $x$ variable (i).
For the Blasius and Langford systems, gray boxes [in (f) and (i)] indicate regions of approximately monotonic scaling, between the parameter and statistical feature, in which \ourmethod{} succeeds [as in (e) and (h)].
The vertical dotted lines in (c), (f), and (i) mark the default parameter value used in the simulation of each system.}
\end{figure*}

\subsection{\label{experiment_amplitude}Parameter amplitude}

In addition to parameter timescale, the amplitude of parameter variation is another potential determinant of the performance of \problemname{} algorithms.
For a given process, a TVP $\theta_t$ with very small parameter amplitude may produce negligible fluctuations in the joint probability distribution $p(x_t,..., x_{t+W}|\theta_t)$, and hence statistics $f:(x_t,...,x_{t+W}) \rightarrow \mathbb{R}$ of the process.
In the limiting case, as the amplitude of parameter variation goes to zero, the process approaches stationarity.
\problemname{} may fail in such low-TVP-amplitude cases due to the challenge of detecting increasingly subtle dynamical variation.
Additionally, a very large parameter amplitude can produce qualitative changes in system behavior, such as bifurcations, that disrupt a monotonic relationship between a TVP and the dynamical properties of the process, making statistical inference of an underlying TVP more challenging in this setting also.
Further subtleties that could impact performance include the position of a system parameter with respect to a bifurcation (i.e., very close versus very far) and the magnitude of dynamical variation relative to the magnitude of the TVP.

To investigate the dependence of \ourmethod{} performance on parameter amplitude, we simulated non-stationary Lorenz, Blasius, and Langford processes driven by slow sinusoidal TVPs with parameter amplitudes ranging over up to seven orders of magnitude relative to the baseline value.
Results are shown in Fig.~\ref{fig_amplitude}.
For each process, the performance of \ourmethod{} is plotted for different proportional amplitudes of parameter variation [Figs~\ref{fig_amplitude}(b), (e), and (h)], corresponding to different values of $\alpha$ in Eq.~\eqref{eq_tvp}, which specifies the maximum magnitude of variation of each TVP around the default value for each system (defined in Appendix~\ref{appendix}).
\ourmethod{} achieves near-perfect performance ($R^2 > 0.99$) over multiple orders of magnitude for each system.
As expected, for all three systems, \ourmethod{} fails at very low amplitudes of parameter variation, corresponding to regimes in which the effect of parameter variation on the statistics of dynamical properties of each process is too small to be detected.
Notably, as the amplitude increases, the performance deteriorates again for both the Blasius and Langford processes, whereas an upper limit on \ourmethod{} performance is not seen for the Lorenz process.

To better understand why the performance of \ourmethod{} deteriorates with increasing parameter amplitude for some processes but not others, we examined the behavior of statistics from the \textit{catch24} feature set over a range of parameter values for each process.
We present an exemplar statistic for each process that helps to explain the performance of \ourmethod{} relative to parameter amplitude in Figs~\ref{fig_amplitude}(c), (f), and (i).
For the Lorenz process, the mean value of the variable $z$ increases approximately monotonically with $\rho$ over a wide range of considered amplitudes [Fig.~\ref{fig_amplitude}(c)] and \ourmethod{} succeeds over this same range [Fig.~\ref{fig_amplitude}(b)].
In contrast, for both the Blasius and Langford processes, there exists some amplitude of parameter variation beyond which \ourmethod{} fails.
In both cases this can be explained by the existence of a bounded region of monotonic scaling (denoted with a gray rectangle) between the parameter and the feature, centered around the default parameter value (denoted with a dotted line) [Fig.~\ref{fig_amplitude}(f), (i)].
For the Blasius process, over a small range of the parameter $\alpha_1$, there is an approximately monotonic relationship between the parameter and the \textit{catch22} feature \texttt{stretch\_decreasing} applied to variable $x$ [Fig.~\ref{fig_amplitude}(f)]; this function computes the longest decreasing sequence of time-series values.
As the value of $\alpha_1$ decreases $\lessapprox 0.186$, the $ax$ term comes to dominate the $\dot x$ equation of the Blasius ODE (Eq.~\eqref{eq_Blasius}) and so the $x$ variable (which represents the amount of vegetation in the Blasius foodweb model) increases exponentially, and hence \texttt{stretch\_decreasing} is equal to zero for these values.
Hence \texttt{stretch\_decreasing} cannot be used to track an $\alpha_1$ TVP when its values fall below this threshold.
For the Langford process, the value of the feature \texttt{forecast\_error} for variable $x$ scales approximately linearly with $|\omega|$ [Fig.~\ref{fig_amplitude}(i)]; noting that this \texttt{forecast\_error} returns the standard deviation of the residuals of a sliding forecast based on the mean of the preceding three time-series values.
For values of $\omega$ close to zero, the Langford process slows down, so these simple forecasts become increasingly accurate, and the feature \texttt{forecast\_error} reaches a local minimum.
Consequently, \texttt{forecast\_error} can be used to approximate the time variation of $\omega$ provided that its values remain within one of these two monotonic scaling regions.
In short, for the Blasius and Langford processes, the upper limit of \ourmethod{} performance with respect to proportional amplitude variation is determined by the position of the default parameter value within a region of monotonic scaling between the parameter and at least one statistic in the \textit{catch24} feature set.


In summary, the success of \ourmethod{} requires the amplitude of variation of the TVP of interest to be large enough to produce non-stationarity on timescales of interest and small enough that an approximately monotonic relationship between the TVP and certain dynamical properties is maintained.
More generally, we expect these considerations regarding parameter amplitude to apply to the broader class of \problemname{} algorithms.
\begin{figure*}
\includegraphics[width=350pt]{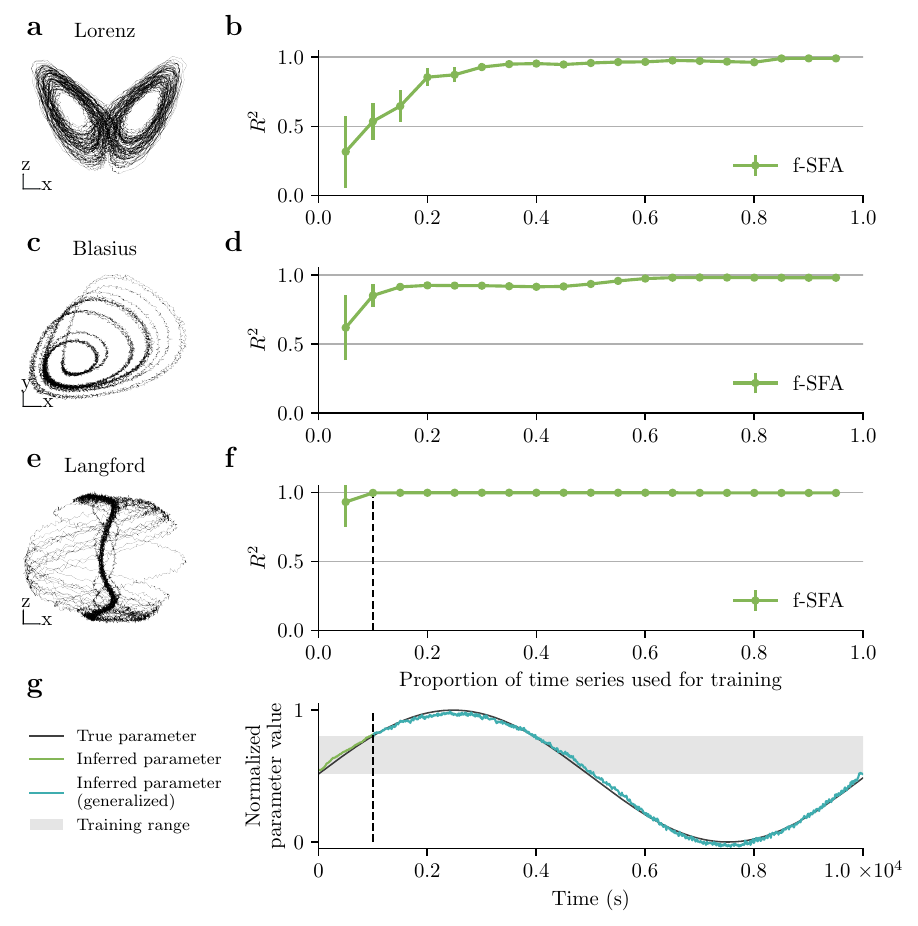}
\caption{\label{fig_generalisation}
\textbf{\ourmethod{} can generalize to unseen parameter values:}
Non-stationary Lorenz [(a)-(b)], Blasius [(c)-(d)], and Langford [(e)-(f)] processes were simulated using a slow sinusoidal TVP and SNR 20\,dB in each case.
Visualizations [(a), (c), (e)] depict 2-dimensional projections of the noisy processes.
Plots of Pearson $R^2$ \problemname{} performance [(b), (d), (e)] show the correlation between the inferred and true TVP across the entire time series after training \ourmethod{} with initial time-series segments of varying proportion.
A TVP reconstruction example is shown in (g) corresponding to the 0.1 proportion case indicated by a dotted line for the Langford system (f), and with the gray shaded region corresponding to the range of TVP values seen in the training data.}
\end{figure*}

\subsection{\label{experiment_ts_length}Generalization to unseen parameter values}

Since \ourmethod{} produces a linear transformation that operates on a statistical feature space, it is possible to use training data to learn a \ourmethod{} transformation which is then applied to a feature-space embedding of new time-series data.
For example, one could train a \ourmethod{} model for the purpose of tracking parameter variation as part of an industrial process, which is later used for online monitoring of the process.
But is strong performance restricted to the parameter ranges on which \ourmethod{} is applied, or can it generalize to unseen parameter values?
To explore this problem, we examined whether \ourmethod{} applied to an initial segment of a time series can successfully generalize to TVP inference across the entire time series.

To investigate the possibility of generalizing \ourmethod{} to unseen parameters, we simulated the same three chaotic flows used in Sec.~\ref{experiment_comparative}: the Lorenz, Blasius, and Langford processes.
In each case, the TVP was a single-period sinusoid with an amplitude of $\pm 10\%$ of the value of the default parameter.
Gaussian measurement noise with SNR of 20\,dB was added to each time series.
For each time series of length $T$, \ourmethod{} was applied to an initial time-series segment of length $\lfloor \beta T \rfloor$ samples, $\beta \in (0,1)$, and then the linear transformation learned with \ourmethod{} was applied to infer a TVP for the remaining segment of length $\lfloor(1 - \beta)T\rfloor$ samples, for an ascending sequence of $\beta$.
Considering the functional form of the sinusoid, the proportion of the total parameter range used to learn the linear transformation was $29\%$ for $\beta = 0.1$ and $50\%$ for $\beta \in [0.25, 0.5]$.
Although this is an unsupervised learning problem, in this problem we refer to the initial time-series segment as the training segment and the latter segment as the generalization, or test, segment.

As shown in Fig.~\ref{fig_generalisation}, generalization of \ourmethod{} to unseen parameter values was possible for all three systems.
Since the TVP was a slow sinusoid in each case, the TVP ranges for the first and second halves of the time series were disjoint.
Consequently, good \ourmethod{} performance when trained with $<50\%$ of the time-series samples (hence less than half the range of parameter values) clearly indicates a successful generalization to unseen parameter values [Figs.~\ref{fig_generalisation}(b), (d), (f)].
Indeed, generalization succeeded for the Blasius and Langford systems with as few as $10\%$ of the samples for each time series (corresponding to $29\%$ of the parameter range).
An example TVP inferred via \ourmethod{} generalization is shown for the Langford process in Fig.~\ref{fig_generalisation}(g).
Taking into account the results of Sec.~\ref{experiment_amplitude} (see Fig.~\ref{fig_amplitude}), generalization succeeds in these case studies because \ourmethod{} is learning a linear feature-to-parameter mapping over a small range of parameter values which captures the (approximately linear) monotonic scaling between feature and parameter values over a larger range.
Indeed, whenever an approximately linear relationship exists between features and a parameter values over a suitably large range, within which \ourmethod{} is applied, generalization to unseen values within this range will be possible.
\section{\label{sleep_application}Application: tracking sleep depth}

\begin{figure*}
\includegraphics[width=500pt]{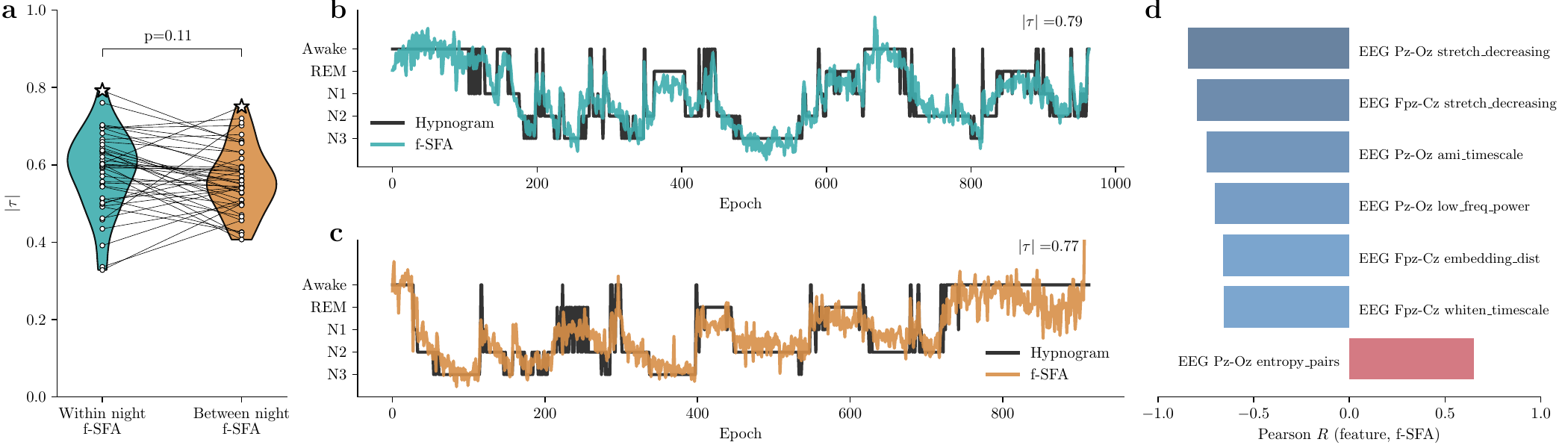}
\caption{\label{fig_sleep_depth}
\textbf{Applying \ourmethod{} to polysomnography (PSG) data yields a time-varying parameter that tracks sleep depth:} \ourmethod{} using \textit{catch22} features was applied to overnight PSG data from the Sleep-EDF dataset \cite{kempAnalysisSleepdependentNeuronal2000}
(a) A violin plot of the Kendall's tau correlation $|\tau|$ between the hypnogram and the most correlated of the first two SFA components from \ourmethod{} (taking the best performance across the two components) for each of the Sleep-EDF PSG recordings.
(b) The hypnogram (in black) and first \ourmethod{} component (in teal) for a within-night application of our method to an exemplar PSG [marked with a star in panel (a)].
(c) The hypnogram (in black) and first \ourmethod{} component (in orange) for a between-night application of our method to an exemplar PSG [marked with a star in panel (a)].
(d) A bar plot of Pearson correlation (negative in blue and positive in red) between channel--feature pairs and the inferred TVP in panel (b), featuring the subset of channel--feature pairs with the largest absolute correlation.
}
\end{figure*}


Non-stationarity is evident in electroencephalography (EEG) and other recordings obtained during human sleep, with slow variation in statistical time-series properties seen between and within different stages of sleep~\cite{metznerSleepRandomWalk2021, tamakiSurveillanceREMSleep2019}.
Accordingly, in this section we aimed to test a correspondence between the slow variation inferred from data (via \ourmethod{}), with the traditionally manual annotations of sleep into five stages: wake, rapid eye movement (REM), and non-REM stages N1, N2, and N3~\cite{berry2012aasm}.
This expert-rated sequence of sleep stages across non-overlapping 30\,s windows (termed epochs), is called sleep staging, with the resulting chart called a hypnogram.
The problems with conventional sleep staging are well-known.
First, manual sleep scoring is expensive, and yields sleep-stage assignments that vary substantially between scorers (e.g., a reported Cohen's kappa inter-rater reliability of 0.76 \cite{leeInterraterReliabilitySleep2022}).
Furthermore, there is considerable dynamical heterogeneity within sleep stages, indicating that the existing sleep-staging system oversimplifies the subtle and complex dynamical variation observed across a sleep episode \cite{decatTraditionalSleepScoring2022a, tamakiSurveillanceREMSleep2019}.
An alternative approach to sleep-stage discretization is to infer a low-dimensional continuous representation of sleep dynamics, e.g., via dimension reduction or techniques such as self-supervised learning \cite{mcphersonCharacterizationSleepUsing2001, wuAssessSleepStage2015, katzAlternatingDiffusionMaps2019,
liuExploreIntrinsicGeometry2021, banvilleUncoveringStructureClinical2021, yeCoSleepMultiViewRepresentation2022, yangSelfSupervisedElectroencephalogramRepresentation2023, metznerExtractingContinuousSleep2023}.
Assuming that the hypnogram tracks the dominant axis of slow, non-stationary variation during sleep, we hypothesized that the TVP inferred by applying \ourmethod{} to sleep data would exhibit a significant correlation with the corresponding hypnogram sequence.
Such a correlation would indicate that the unsupervised data-driven inference of a slow axis of dynamical variation across a sleep is capturing some of the variation captured by the existing sleep stages.
However, we stress that, in the absence of a ground truth, a higher correlation with the manually annotated categorical hypnogram does not strictly indicate better performance; a lower correlation would provide room for the inferred \ourmethod{} component to capture additional (and perhaps complementary) information about sleep dynamics that are not captured by current sleep-staging rules.

\subsection{\label{sec:sleep_methods}Data and methods}

To obtain \ourmethod{} components we used the sleep telemetry subset of the Sleep-EDF dataset \cite{kempAnalysisSleepdependentNeuronal2000} from the PhysioNet database \cite{goldbergerPhysioBankPhysioToolkitPhysioNet2000}.
The dataset comprises 44 polysomnogram (PSG) recordings [22 subjects (7 male), aged 18--79 years, two nights each], each with four channels recorded at a sampling rate of 100\,Hz: a FPz-Cz EEG, a Pz-Oz EEG, a horizontal electrooculogram (EOG), and a submental electromyogram (EMG).
We applied \ourmethod{} to each PSG: first computing a \textit{catch22} feature embedding \cite{lubbaCatch22CAnonicalTimeseries2019} using non-overlapping 30\,s windows that corresponded to the PSG epochs that had been used for sleep-stage assignment by expert raters, and then applying SFA.
We used \textit{catch22} (rather than \textit{catch24}, i.e., not adding mean and standard deviation features) for the feature set because mean drifts in an EEG typically represent undesirable artifact, e.g., due to loss of electrode contact due to perspiration, especially in long records such as PSGs \cite{uriguenEEGArtifactRemoval2015a}.
We applied \ourmethod{} both \textit{within} nights, training and inferring a TVP using a single PSG for a single subject, and \textit{between} nights, training on one PSG and generalizing to another PSG for a single subject.
We included the between-night case to see whether we could potentially learn a good mapping to a driving TVP that generalizes to unseen data.
The sleep-stage categories were wake, REM, N1, N2, and N3 (note that we recoded N4 labels as N3, consistent with the current American Academy of Sleep Medicine (AASM) nomenclature \cite{berry2012aasm}).


To quantify the correlation between the low-dimensional components inferred using \ourmethod{} and the hypnogram time series we used absolute Kendall's tau, $|\tau|$~\cite{kendallNewMeasureRank1938}, which can accommodate ordinal data with repeated values.
For the computation of $|\tau|$, we treated the sleep stages as ordinal values along a single dimension with $\text{wake} = 1$, $\text{REM} = 0$, $\text{N1} = -1$, $\text{N2} = -2$, and $\text{N3} = -3$, reflecting the ordering that is conventionally used in hypnogram plotting.
The ordering from wake to N1--N3 is thought to reflect sleep depth and is supported by an increasing arousal threshold along this axis \cite{rechtschaffenAuditoryAwakeningThresholds1966, bonnetThresholdSleepPerception1982}.

\subsection{\label{sec:sleep_results}Results}

We examined the strength of the correlation between the hypnogram and the first component of \ourmethod{} for each subject.
The distribution of $|\tau|$ values across the 44 PSG recordings is shown in Fig.~\ref{fig_sleep_depth}(a).
For within-night \ourmethod{} (i.e., applying \ourmethod{} to PSG data for a single night and comparing the inferred TVP to the corresponding hypnogram), the mean value of $|\tau|$ was 0.58 and the range was 0.33--0.79, and for between-night \ourmethod{} (i.e., applying \ourmethod{} to PSG data from one night and then generalizing it to unseen PSG data from a second night for the same subject) the mean was 0.55 and the range was 0.40--0.75, consistent with a moderate rank correlation between the inferred TVPs and the hypnograms.
The same subjects were used for the inference within and between nights, so a paired $t$-test was used to compare the means of the two experiments, finding no statistically significant difference between the distributions of the values of within- and between-night $|\tau|$ ($t=1.62$, $\text{df}=43$, $p=0.11$).
This suggests that \ourmethod{} can generalize to unseen PSG data for a given subject. 
The results support our hypothesis that \ourmethod{} can capture a meaningful continuous axis of slow dynamical variation across a sleep recording that recapitulates much of the variation contained in manual sleep staging, while also providing a richer continuous representation (containing information about the dynamical fluctuations in sleep beyond what is contained in the categorical hypnogram).

To demonstrate the level of correspondence between the \ourmethod{} component that could be achieved with the manually scored hypnogram, the best-performing within-night TVP ($|\tau| = 0.79)$ is shown in Fig.~\ref{fig_sleep_depth}(b), and the best performing between-night TVP ($|\tau| = 0.77$) is shown in Fig.~\ref{fig_sleep_depth}(c).
Strikingly, we find that the inferred TVPs follow the overall pattern of each hypnogram and that many of the brief wake periods are matched by short, often single-epoch, upward deflections of the inferred parameters.
Notably, the inferred TVPs provide a \textit{continuous} representation of sleep depth, in contrast to the categorical AASM staging system.
Starting with no knowledge other than trying to find a slow TVP, the unsupervised application of \ourmethod{} to PSG data produces a data-driven component that is significantly correlated with the manually annotated hypnogram.
Since REM fits imperfectly into the sleep-depth axis, from wake through to the progressively deeper NREM stages (N1, N2, N3), we conducted a sensitivity analysis on our results in which we omitted all REM epochs from the analysis.
A paired $t$-test found no statistically significant change in Kendall's tau, with or without REM, for within-night ($t=0.68$, $p=0.50$, df$=43$) or between-night ($t=0.73$, $p=0.47$, df$=43$) TVPs.

Next, we sought to understand how the first \ourmethod{} components track the hypnogram by leveraging the interpretability of the \textit{catch22} time-series features.
Recall that each \ourmethod{} component is obtained via a linear transformation of the time series of features in the matrix $F$ (Eq.~\eqref{eq_F}), allowing us to determine which time-series features applied to which variables contribute most to a given component.
For the within-night TVP shown in Fig.~\ref{fig_sleep_depth}(b), we computed Pearson correlations $R$ between the TVP and each of the individual \textit{catch22} feature time series, of which the largest absolute correlations are shown in Fig.~\ref{fig_sleep_depth}(d).
The largest positive correlation $R$ was seen for \texttt{entropy\_pairs} applied to the Pz-Oz EEG channel, consistent with the previous finding of increased entropy in wake and REM states compared to NREM sleep~\cite{buriokaApproximateEntropyElectroencephalogram2005}.
The largest negative correlations were for the length of stretches of monotonically decreasing time-series values (\texttt{stretch\_decreasing}), the first minimum of the automutual information function (\texttt{ami\_timescale}), and increased low-frequency power (\texttt{low\_freq\_power}), consistent with the NREM predominance of lower frequency bands, notably the delta oscillations (0.5--4~Hz) of `slow-wave sleep' typical of N3, which are characterized by higher autocorrelation, dominant low frequencies, and longer monotonic segments compared to wake and REM stages.
We can interpret the \ourmethod{} component as approximately varying along a continuous dimension of sleep depth, ranging from the spontaneous, complex activity of the wake and REM states at one end, to the slower and more periodic activity of N3 at the other.

In summary, \ourmethod{} applied to PSG data yields a component that captures the slow-varying statistical properties of sleep that is associated with variation in sleep depth, supporting the use of \ourmethod{} as an unsupervised, data-driven method that is well-suited to discovering hidden non-stationary variation in complex empirical recordings.



\section{\label{discussion}Discussion}

In this paper we introduced \ourmethod{}, an unsupervised data-driven method for tracking non-stationary statistical variation across time-series data that is sensitive to a wide range of potential sources of variation through the use of a diverse time-series feature set (such as \textit{catch22} \cite{lubbaCatch22CAnonicalTimeseries2019}).
Our method bypasses the need for manual selection of relevant time-series statistics that correlate with the TVP~\cite{guttlerReconstructionParameterSpaces2001}, demonstrating that non-stationary variation can be inferred from time-series data using the combination of:
(i) a comprehensive set of candidate time-series statistics (to be sensitive to a wide range of potential sources of dynamical variation); and
(ii) the slowness inductive bias of SFA to isolate a candidate prediction for the TVP as a source of slow statistical variation on a given timescale (that is highly robust to irrelevant features, unlike variance-based dimension-reduction methods, cf. Sec.~\ref{feature_feature}).
The performance and noise robustness of \ourmethod{} using \textit{catch24} features substantially outperformed that of a representative set of comparator \problemname{} methods across a range of non-stationary processes (Sec.~\ref{experiment_comparative}), and demonstrated that \ourmethod{} using only mean and variance serves as a surprisingly strong baseline method that could be used as a comparator in the future development of new \problemname{} algorithms.
We then investigated key determinants of \ourmethod{} performance, finding that parameter inference is more likely to succeed when the timescale of parameter variation is sufficiently slow relative to the windows used for computing time-series statistics (Sec.~\ref{experiment_timescale}), and when an approximately monotonic relationship is obtained between some time-series feature and the underlying TVP  (Sec.~\ref{experiment_amplitude}).
Through synthetic case studies, we showed that \ourmethod{} can generalize to unseen parameter values for some processes (Sec.~\ref{experiment_amplitude}).
Additionally, applying \ourmethod{} to a sleep dataset yields a measure of slow, non-stationary variation that tracks sleep depth and correlates with expensive, manually annotated categorical scoring of the dataset by sleep experts, while providing a dynamically richer continuous representation of the dynamics (Sec.~\ref{sleep_application}).
This ability to track meaningful variation hidden in complex empirical time-series data provides evidence that \ourmethod{} has real-world utility beyond tracking parameter variation in simulated chaotic systems.

Using a comparative methodology to benchmark algorithms across different non-stationary processes is essential to advance research on \problemname{} methods.
Historically, such comparison has been lacking due to the disjoint nature of the literature, with publications appearing in different academic fields and with limited referencing between disciplines.
In \citet{owensParameterInferenceNonstationary2024}, we addressed this fragmentation by defining \problemname{} and organizing the literature into different categories of methods.
We also observed that most previous work had examined only the performance of a single algorithm on one or two handpicked systems, such as the Lorenz process or the logistic map (which allow trivial tracking through windowed mean and variance, respectively) \cite{owensParameterInferenceNonstationary2024}.
In contrast, and to our knowledge, the present work is the first attempt to compare a range of \problemname{} methods across different systems and noise levels.
Such comparisons are necessary to understand how any new algorithm performs and relates to the state of the field.
However, a major remaining barrier to comparative \problemname{} research is that most published algorithms lack open implementations.
We have taken steps towards addressing this by making available our implementation of \ourmethod{} and the various experiments in this paper, and we hope that future algorithm development will build on our research by using \ourmethod{} as a comparator.

\ourmethod{} provides a data-driven solution to the problem of selecting appropriate statistical features for a given problem.
Previously, the question of how to optimally choose a set of time-series features for use in \problemname{} had not been resolved, that is, for the category of methods that we call `statistical times-series feature-based methods' \cite{owensParameterInferenceNonstationary2024}.
\citet{guttlerReconstructionParameterSpaces2001} demonstrated that there exist time-series features that track parameter variation for certain processes, but left the issue of feature selection as an open problem.
Given a set of time-series features, \citet{chatterjeeOptimalTrackingParameter2002} later showed that applying smooth orthogonal decomposition (which is equivalent to SFA) to a multivariate time series of such features provides optimal tracking of an underlying TVP for chaotic systems, in an asymptotic sense relative to time-series length, but their choice of features remained arbitrary.
By construction, general purpose feature sets (for which \textit{catch24} is a compact example) are sensitive to a wide range of dynamical properties, and thus offer a good starting point for \ourmethod{}, and feature-based \problemname{} more generally.
Alternative ways to select features include using existing time-series feature sets that are specialized for a particular domain (e.g., for sleep EEG \cite{cabralFATSFeetsFurther2018} or function magnetic resonance imaging \cite{alamCanonicalTimeseriesFeatures2024}), or by selecting high-performing features from a comprehensive set like \textit{hctsa} when time series are available for which the ground truth TVP is known.
Moreover, we demonstrated that SFA does not suffer from PCA's bias towards TVP-uninformative features that contribute a high amount of explained variance through their covariation (due to the construction of a feature set that includes algorithmic redundancy) (see Sec.~\ref{feature_feature}).
Accordingly, \ourmethod{} could be performed using more comprehensive feature sets such as \textit{hctsa} \cite{fulcherHctsaComputationalFramework2017} (or \textit{pyhctsa} \cite{moorePyhctsaPythonPackage2026}) or the features packaged in \textit{theft} \cite{hendersonFeatureBasedTimeSeriesAnalysis2025}.
In principle, such large feature sets with a greater coverage of more complex and subtle time-series properties could be even more sensitive to different sources of slow non-stationary statistical variation than we report for \ourmethod{} using \textit{catch24}, but at the cost of a greater computational expense (that could be partially offset through the parallelization of feature computation).
Further, if feature interpretability is prioritized, then \ourmethod{} can instead be performed using simpler feature sets such as distributional moments, distributional centiles, or spectral power.
In particular, our results support this use of \baseline{} as an interpretable baseline, in keeping with previous observations that mean and variance perform surprisingly well on some \problemname{} problems \cite{owensParameterInferenceNonstationary2024} as well as time-series classification tasks \cite{hendersonNeverDullMoment2023}.
Indeed, the strikingly poor performance of several sophisticated \problemname{} algorithms compared to this baseline in Sec.~\ref{experiment_comparative} highlights the importance of examining data for time-varying distributional moments.

We have argued that timescale separation is a key determinant of \ourmethod{} performance (in Sec.~\ref{experiment_timescale}).
Indeed, the same considerations presumably apply to window-based \problemname{} methods in general, in that the window timescale must be long enough to reliably estimate a statistic and short enough that the TVP is approximately stationary over that time period.
Such methods are best suited to inferring TVPs that are slow relative to the observed data, and will also tend to require long time series with high sampling rates.
Indeed, this is precisely why we applied \ourmethod{} to the PSG data in Sec.~\ref{sleep_application}, because whole-night neural electrophysiology data satisfy these requirements (e.g., with an 8-hour, 100\,Hz EEG recording yielding 960 windows of length 30 seconds, with 3000 samples per window).
Thus, the first key limitation of \ourmethod{} is that the TVP must be sufficiently slow, otherwise the inductive bias of SFA towards slowness will lead to poor performance.
If the TVP is not slow, alternative \problemname{} methods need to be considered, such as those that transform the raw time series (e.g., quadratic SFA \cite{wiskottEstimatingDrivingForces2003}), or else a supervised learning paradigm may be required, starting with data for which a TVP is known and then generalizing to time series outside the training set.
The second key limitation of \ourmethod{} is the requirement that the selected set of time-series features contain at least one statistic that is sensitive to the non-stationary variation of the observed process.
In part, this issue can be addressed by using a comprehensive set of time-series features when applying \ourmethod{}, but, as noted above, this entails a trade-off with computational cost.


Several avenues for further research warrant mention.
First, the choice of window length in \ourmethod{} could be optimized by an algorithmic determination of the timescale of the process, or by searching across values of $W$ (and/or $S$) for which \ourmethod{} performance converges in the hyperparameter space.
Second, the performance of alternative dimension-reduction methods could be compared to that of SFA.
Indeed, elsewhere we have demonstrated that SFA is mathematically related to a range of time-series dimension reduction (TSDR) algorithms that extract components based on temporal properties such as autocorrelation and predictability \cite{owensTimeseriesDimensionReduction}, most of which could be used to construct algorithmic variants of \ourmethod{}.
Third, as noted in Appendix~\ref{appendix_multiple}, inferring multiple parameters using \ourmethod{} is possible for the Lorenz process, at least in some cases where there is timescale separation between each TVP, allowing them to be distinguished and ordered by SFA according to `slowness'.
However, in general we expect \problemname{} with multiple TVPs to be challenging.
Further work is needed to optimize \ourmethod{} for this setting.
Finally, although we focused on univariate statistics in this work, more generally, \ourmethod{} could be performed using pairwise \cite{cliffUnifyingPairwiseInteractions2023a} or even multivariate (e.g., graph-based \cite{peachHCGAHighlyComparative2021}) time-series feature sets.
One can conceive of scenarios where a TVP modulates pairwise relationships between variables, e.g., through modulation of connectivity in a biological neural networks, leading to changes in pairwise relationships between time series recorded from different areas.
Such extensions may further enhance the sensitivity of \ourmethod{} to non-stationary statistical variations in complex high-dimensional time-series datasets.

In conclusion, \ourmethod{} is a high-performing, noise-robust approach to \problemname{}, that enables the inference of slow time-varying parameters from long, non-stationary time series.
We anticipate that \ourmethod{} will be of value in a range of settings, as diverse as systems biology, physics, and finance, where \ourmethod{} can be used as part of the quantitative analysis of observed time series.


\begin{acknowledgments}
B.D.F. acknowledges support from the Australian Research Council (FT240100418).
\end{acknowledgments}


\appendix
\section{non-stationary processes}
\label{appendix}

Here we define the chaotic maps and flows that were used in our numerical experiments.
In each case, we specify the system equations, initial condition, and parameter values.
For details regarding parameter variation and the integration of ODEs see Sec.~\ref{sec_nonstat_procs}.

\subsection{The Logistic Map}

For the logistic map
\cite{maySimpleMathematicalModels1976} with equation:
\begin{align}
    \label{eq_logistic}
    x_{n+1} &= r x_n (1 - x_n)\,,
\end{align}
we used a parameter value of $r = 3.6$ and the initial condition $x_0 = 0.6$.

\subsection{The Sine Map}

For the sine map
\cite{sprottChaosTimeSeriesAnalysis2001} with equation:
\begin{align}
    \label{eq_sine}
    x_{n+1} &= r \sin(\pi x_n)\,,
\end{align}
we used a parameter value of $r = 3.0$ and the initial condition $x_0 = 0.6$.

\subsection{The predator--prey map}

The predator--prey map models the population dynamics of two species with non-overlapping generations
\cite{beddingtonDynamicComplexityPredatorprey1975}.
It is defined as:
\begin{subequations}
\label{eq_pred_prey}
\begin{align}
    x_{n+1} &= x_n \exp (r(1 - x_n/K) - ay_n)\,, \\
    y_{n+1} &= ax_n (1 - \exp (-ay_n))\,,
\end{align}
\end{subequations}
where parameters were set to values $r=3.0$, $K=1.0$, and $a=5.0$, and we used the initial condition $(x_0,y_0)= (0.5, 0.5)$.

\subsection{The Lorenz process}

The Lorenz process is a model of atmospheric convection
\cite{lorenzDeterministicNonperiodicFlow1963} defined by:
\begin{subequations}
\label{eq_lorenz}
\begin{align}
    \dot x &= \sigma(y - x)\,, \\
    \dot y &= x(\rho - z) - y\,, \\
    \dot z &= xy - \beta z\,,
\end{align}
\end{subequations}
where parameters were set to values $\sigma=10$, $\rho=28$, and $\beta = 8/3$ and we used the initial condition $(x,y,z)= (-9.79, -15.04, 20.53)$ from
\citet{gilpinChaosInterpretableBenchmark}.

\subsection{The Langford process}

The Langford process yields a torus-like attractor
\cite{langfordNumericalStudiesTorus1984} defined by:
\begin{subequations}
\label{eq_Langford}
\begin{align}
    \dot x &= (z - \beta)x - \omega y\,, \\
    \dot y &= (z - \beta)y\ + \omega x,, \\
    \dot z &= \lambda + \alpha z - \frac{z^3}{3} - (x^2 + y^2)(1 + \rho z) + \varepsilon z x^3\,,
\end{align}
\end{subequations}
where parameters were set to values $\alpha=0.95, \beta=0.7, \lambda=0.6, \omega=3.5, \rho=0.25, \text{and } \varepsilon=0.1$ and we used the initial condition $(x,y,z)= (-0.78, -0.63, -0.18)$ from \cite{gilpinChaosInterpretableBenchmark}.
Note that the Langford process has also been referred to as the Aizawa process \cite{gilpinChaosInterpretableBenchmark}.

\subsection{The Blasius process}

The Blasius process is a chaotic food web model of vegetation ($x$), herbivores ($y$), and predators ($z$),
\cite{blasiusComplexDynamicsPhase1999} defined by:
\begin{subequations}
\label{eq_Blasius}
\begin{align}
    \dot x &= ax - \frac{\alpha_1 x y}{1 + k_1 x}\,, \\
    \dot y &= -by + \frac{\alpha_1 x y}{1 + k_1x} - \frac{\alpha_2 y z}{1 + k_2 y}\,, \\
    \dot z &= -c(z - z^*) + \frac{\alpha_2 y z}{1 + k_2 y}\,.
\end{align}
\end{subequations}
where parameters were set to $a=1$, $\alpha_1=0.2$, $\alpha_2=1$, $b=1$, $c=10$, $k_1=0.05$, $k_2=0$, and $z^* = 0.006$ and we used the initial condition $(x,y,z) = (4.03, 5.11, 0.0165)$ from \citet{gilpinChaosInterpretableBenchmark}.

\section{\ourmethod{} sensitivity analysis using different time-series features}
\label{appendix_results}

We examined the sensitivity of our results in Sec.~\ref{sec_results} to the choice of time-series features and the dimension-reduction method.
Consider that time-series features can be roughly arranged along an axis from simple (e.g., mean, variance, and distributional quantiles) to complex (e.g., nonlinear correlation structure).
To compare performance across this hierarchy, we repeated the experiment in Sec.~\ref{sec_results} with \ourmethod{} using: (i) twenty evenly spaced quantiles, (ii) time-varying frequency power computed via the short-time Fourier transform (STFT), (iii) \textit{catch24} features, and (iv) all features combined.

\begin{figure*}
\begin{center}
\begin{minipage}[t]{0.48\textwidth}
\includegraphics[width=\linewidth]{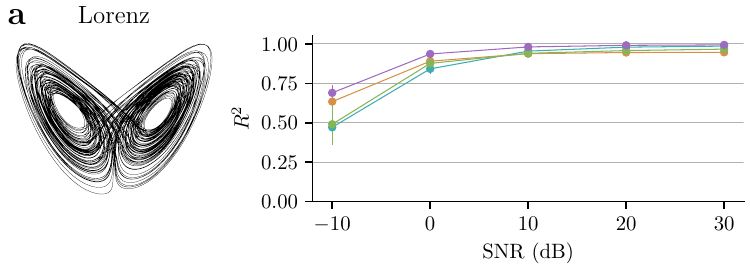}
\end{minipage}\hfill
\begin{minipage}[t]{0.48\textwidth}
\includegraphics[width=\linewidth]{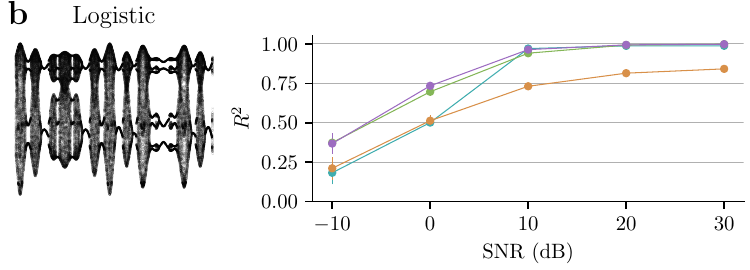}
\end{minipage}
\end{center}
\begin{center}
\begin{minipage}[t]{0.48\textwidth}
\includegraphics[width=\linewidth]{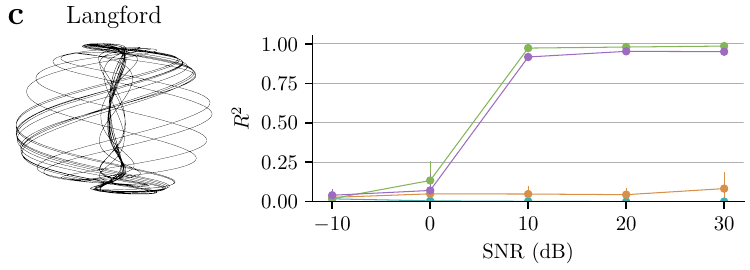}
\end{minipage}\hfill
\begin{minipage}[t]{0.48\textwidth}
\includegraphics[width=\linewidth]{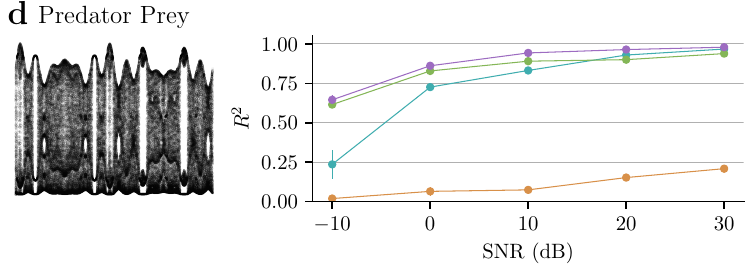}
\end{minipage}
\end{center}
\begin{center}
\begin{minipage}[t]{0.48\textwidth}
\includegraphics[width=\linewidth]{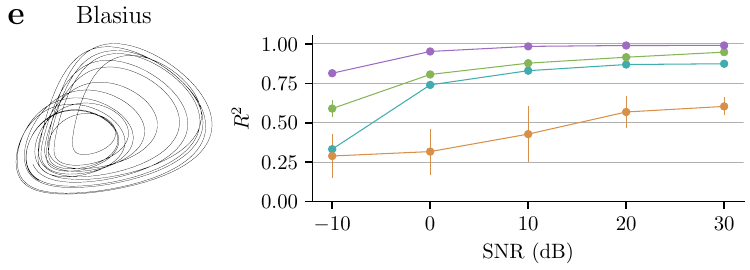}
\end{minipage}\hfill
\begin{minipage}[t]{0.48\textwidth}
\includegraphics[width=\linewidth]{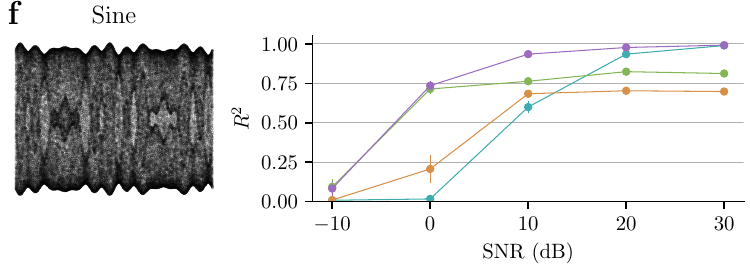}
\end{minipage}
\end{center}
\begin{center}
\includegraphics[width=0.1\textwidth]{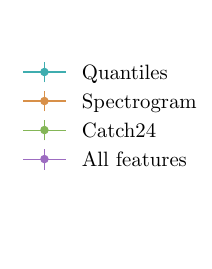}
\end{center}
\caption{\label{fig:sensitivity}
\textbf{A sensitivity analysis of \ourmethod{} to the choice of time-series features finds superior performance with SFA in conjunction with either \textit{catch24} features or a combination of all features.}
The experiment from Sec.~\ref{experiment_comparative} was repeated for \ourmethod{} using different time-series feature sets (i.e., quantiles, spectrogram frequencies, \textit{catch24}, or a combination) to examine the sensitivity of performance to the choice of features.
Each panel visualizes a non-stationary process alongside a plot of \ourmethod{} performance ($R^2$) across different SNRs (dB) for each feature set.
Results are shown for the (a) Lorenz, (b) logistic map, (c) Langford, (d) predator-prey, (e) Blasius, and (f) sine map systems.
}
\end{figure*}

The results of this experiment are shown in Fig.~\ref{fig:sensitivity}.
\ourmethod{} using quantiles performed well for each of the chaotic maps across two or more SNR levels, whereas time-varying frequency power computed via the STFT only performed well for the Lorenz process.
This finding indicates that quantiles and time-varying frequency power offer simple and interpretable alternative feature sets for \ourmethod{} that work for a subset of processes.
In contrast, \ourmethod{} using \textit{catch24} performed well for all systems, supporting the use of this set of features as a computationally efficient default option.
In many cases the performance was similar between \ourmethod{} using \textit{catch24} and all features combined, but with an advantage for the latter seen for the Blasius and sine map processes over several SNR levels.
Accordingly, it may be possible to use larger feature sets to improve performance in some cases.

\section{a case study of inferring multiple time-varying parameters using \ourmethod{}}
\label{appendix_multiple}


\begin{figure*}
\begin{center}
\begin{minipage}[t]{0.3\textwidth}
\includegraphics[width=\linewidth]{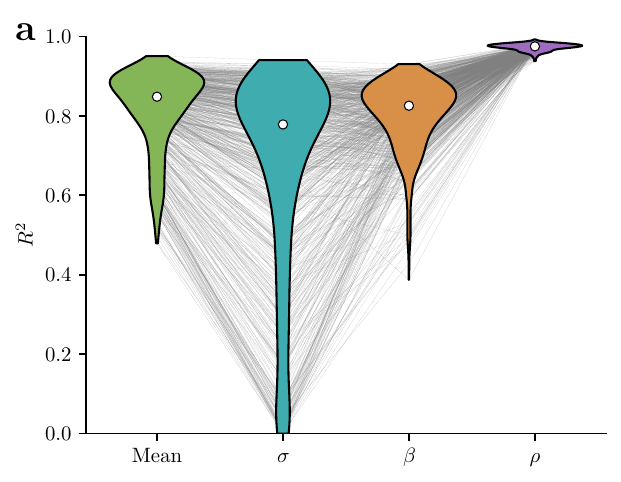}
\end{minipage}\hfill
\begin{minipage}[t]{0.38\textwidth}
\includegraphics[width=\linewidth]{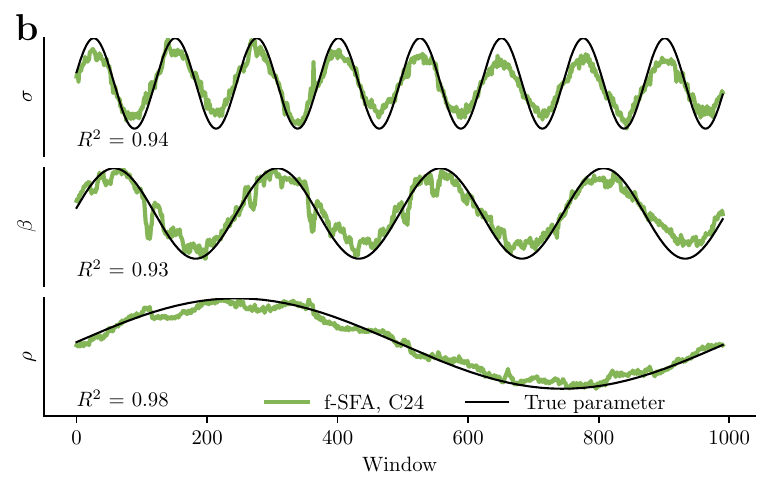}
\end{minipage}\hfill
\begin{minipage}[t]{0.24\textwidth}
\includegraphics[width=\linewidth]{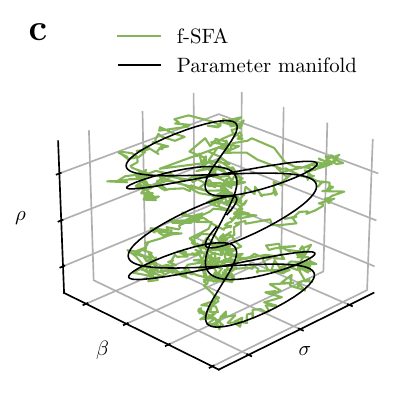}
\end{minipage}
\end{center}
\caption{\label{fig_multi_param}
\textbf{\problemname{} with multiple TVPs is possible for the Lorenz process using \ourmethod{}:}
A series of 720 non-stationary Lorenz processes were simulated that were driven by three sinusoidal TVPs determined by the values of $K_\sigma$, $K_\beta$, and $K_\rho$ in Eq.~\ref{eq_multi_sine} followed by \problemname{} using \ourmethod{}.
(a) Mean performance ($R^2$), and $R^2$ for each TVP, is shown for different integer values of $K_\sigma$, $K_\beta$ and $K_\rho$.
(b) The \ourmethod{} components for which mean $R^2$ was highest are shown (green) along with the corresponding true TVPs (black), with Pearson $R^2$ shown for each pair.
(c) In the setting of multiple TVPs, $f$-SFA can also be conceptualized as inferring a TVP manifold/trajectory. 
}
\end{figure*}

The experiments in our above paper only considered the case where non-stationarity was driven by a single TVP.
Non-stationarity due to multiple TVPs presents a considerably more difficult inference problem.
For example, multiple TVPs could modulate shared set of dynamical properties, making it difficult to isolate the dynamics of each independent TVP.
Additionally, TVPs may affect dynamical properties on different timescales, requiring a multiscale algorithm to infer both parameters, and dynamical variation due to a high-amplitude TVP may mask that of a low-amplitude TVP.
Despite these challenges, \ourmethod{} can in principle infer multiple TVPs in the ideal setting of multiple independently varying parameters that each correspond to unique changes in dynamical properties on a common timescale (as per the ranges shown in Fig.~\ref{fig:timescale}).

To demonstrate the possibility of inferring multiple TVPs using \ourmethod{}, we simulated a series of non-stationary Lorenz processes, the three parameters of which ($\rho$, $\beta$, and $\sigma$) each varied according to a sinusoidal TVP of different frequencies, yielding a time-varying parameter vector, $\boldsymbol{\theta}_t$.
To generate these processes, the TVPs were varied according to slow sinusoids:
\begin{equation}
    \label{eq_multi_sine}
        \sin(2\pi K_pt/T_\text{max}),
\end{equation}
for $p \in \{\rho, \beta, \sigma\}$, using each permutation $(K_\rho, K_\beta, K_\sigma) \in \{1,2,3,\dots,10\}^3$, excluding permutations where $K_\rho=K_\beta$, $K_\rho=K_\sigma$, or $K_\beta=K_\sigma$, to generate a total of 720 non-stationary Lorenz processes.
The amplitude of each parameter was $\pm10\%$ of the default value.
In each case, we applied \ourmethod{} and used the first three components of \ourmethod{} as estimates of the respective TVPs.
To assess the performance of each application of \ourmethod{}, we computed the mean Pearson $R^2$ for each bijection between the TVPs and the three \ourmethod{} components and selected the best value of this.

The results are shown in Fig.~\ref{fig_multi_param}.
\problemname{} with multiple TVPs is possible for the Lorenz process using \ourmethod{} [Fig.~\ref{fig_multi_param}(a)], but, as expected, performance varies considerably depending on the functional form of the TVPs.
However, a good performance of \ourmethod{} (mean $R^2 > 0.8$) was observed for 87\% of the sinusoidal TVP permutations and in all cases mean $R^2 > 0.6$ was achieved.
The best reconstruction (mean $R^2 = 0.97$) is visualized in Fig.~\ref{fig_multi_param}(b).
The inference of $\boldsymbol{\theta}_t$ can also be conceptualized as the reconstruction of a one-dimensional manifold (or trajectory) that is embedded in a higher-dimensional space [Fig.~\ref{fig_multi_param}(c)].
This perspective allows us to visualize and quantify the higher-dimensional parameter dynamics underlying non-stationarity, and may be useful for systems where we expect there to be multiple drivers of non-stationarity, e.g., as in complex biological systems like the brain.
In summary, \problemname{} with multiple TVPs is possible for the Lorenz system using \ourmethod{}.
However, reliably inferring multiple TVPs for a range of systems is a challenging problem that requires further research.





\bibliography{ants}

@article{schreiberClassificationTimeSeries1997,
  title = {Classification of {{Time Series Data}} with {{Nonlinear Similarity Measures}}},
  author = {Schreiber, Thomas and Schmitz, Andreas},
  year = {1997},
  month = aug,
  journal = {Physical Review Letters},
  volume = {79},
  number = {8},
  pages = {1475--1478},
  publisher = {{American Physical Society}},
  doi = {10.1103/PhysRevLett.79.1475},
  urldate = {2023-04-13}
}

@article{guttlerReconstructionParameterSpaces2001,
  title = {Reconstruction of the Parameter Spaces of Dynamical Systems},
  author = {G{\"u}ttler, S. and Kantz, H. and Olbrich, E.},
  year = {2001},
  month = may,
  journal = {Physical Review. E, Statistical, Nonlinear, and Soft Matter Physics},
  volume = {63},
  number = {5 Pt 2},
  pages = {056215},
  issn = {1539-3755},
  doi = {10.1103/PhysRevE.63.056215},
  langid = {english},
  pmid = {11414998}
}

@article{lubbaCatch22CAnonicalTimeseries2019,
  title = {Catch22: {{CAnonical Time-series CHaracteristics}}},
  shorttitle = {Catch22},
  author = {Lubba, Carl H. and Sethi, Sarab S. and Knaute, Philip and Schultz, Simon R. and Fulcher, Ben D. and Jones, Nick S.},
  year = {2019},
  month = nov,
  journal = {Data Mining and Knowledge Discovery},
  volume = {33},
  number = {6},
  pages = {1821--1852},
  issn = {1573-756X},
  doi = {10.1007/s10618-019-00647-x},
  urldate = {2023-09-06},
  langid = {english}
}

@article{fulcherHctsaComputationalFramework2017,
  title = {Hctsa: {{A Computational Framework}} for {{Automated Time-Series Phenotyping Using Massive Feature Extraction}}},
  shorttitle = {Hctsa},
  author = {Fulcher, Ben D. and Jones, Nick S.},
  year = {2017},
  month = nov,
  journal = {Cell Systems},
  volume = {5},
  number = {5},
  pages = {527-531.e3},
  issn = {2405-4712},
  doi = {10.1016/j.cels.2017.10.001},
  urldate = {2023-09-06}
}

@article{gunturkunSequentialReconstructionDrivingforces2010,
  title = {Sequential Reconstruction of Driving-Forces from Nonlinear Nonstationary Dynamics},
  author = {G{\"u}nt{\"u}rk{\"u}n, Ula{\c s}},
  year = {2010},
  month = jul,
  journal = {Physica D: Nonlinear Phenomena},
  volume = {239},
  number = {13},
  pages = {1095--1107},
  issn = {01672789},
  doi = {10.1016/j.physd.2010.02.014},
  urldate = {2023-05-15},
  langid = {english}
}

@article{chatterjeeOptimalTrackingParameter2002,
  title = {Optimal Tracking of Parameter Drif in a Chaotic System: Experiment and Theory},
  shorttitle = {{{OPTIMAL TRACKING OF PARAMETER DRIFT IN A CHAOTIC SYSTEM}}},
  author = {Chatterjee, A. and Cusumano, J.P. and Chelidze, D.},
  year = {2002},
  month = mar,
  journal = {Journal of Sound and Vibration},
  volume = {250},
  number = {5},
  pages = {877--901},
  issn = {0022460X},
  doi = {10.1006/jsvi.2001.3963},
  urldate = {2023-07-22},
  langid = {english}
}

@misc{wiskottEstimatingDrivingForces2003,
  title = {Estimating {{Driving Forces}} of {{Nonstationary Time Series}} with {{Slow Feature Analysis}}},
  author = {Wiskott, Laurenz},
  year = {2003},
  month = dec,
  number = {arXiv:cond-mat/0312317},
  eprint = {cond-mat/0312317},
  publisher = {{arXiv}},
  doi = {10.48550/arXiv.cond-mat/0312317},
  urldate = {2023-04-05},
  archiveprefix = {arxiv}
}

@article{wiskottSlowFeatureAnalysis2002,
  title = {Slow {{Feature Analysis}}: {{Unsupervised Learning}} of {{Invariances}}},
  shorttitle = {Slow {{Feature Analysis}}},
  author = {Wiskott, Laurenz and Sejnowski, Terrence J.},
  year = {2002},
  month = apr,
  journal = {Neural Computation},
  volume = {14},
  number = {4},
  pages = {715--770},
  issn = {0899-7667, 1530-888X},
  doi = {10.1162/089976602317318938},
  urldate = {2023-04-13},
  langid = {english}
}

@article{maySimpleMathematicalModels1976,
  title = {Simple Mathematical Models with Very Complicated Dynamics},
  author = {May, Robert M.},
  year = {1976},
  month = jun,
  journal = {Nature},
  volume = {261},
  number = {5560},
  pages = {459--467},
  publisher = {{Nature Publishing Group}},
  issn = {1476-4687},
  doi = {10.1038/261459a0},
  urldate = {2023-09-30},
  copyright = {1976 Springer Nature Limited},
  langid = {english}
}

@article{beddingtonDynamicComplexityPredatorprey1975,
  title = {Dynamic Complexity in Predator-Prey Models Framed in Difference Equations},
  author = {Beddington, J. R. and Free, C. A. and Lawton, J. H.},
  year = {1975},
  month = may,
  journal = {Nature},
  volume = {255},
  number = {5503},
  pages = {58--60},
  issn = {0028-0836, 1476-4687},
  doi = {10.1038/255058a0},
  urldate = {2023-08-17},
  langid = {english}
}

@article{lorenzDeterministicNonperiodicFlow1963,
  title = {Deterministic {{Nonperiodic Flow}}},
  author = {Lorenz, Edward N.},
  year = {1963},
  month = mar,
  journal = {Journal of the Atmospheric Sciences},
  volume = {20},
  number = {2},
  pages = {130--141},
  publisher = {{American Meteorological Society}},
  issn = {0022-4928, 1520-0469},
  doi = {10.1175/1520-0469(1963)020<0130:DNF>2.0.CO;2},
  urldate = {2023-09-30},
  chapter = {Journal of the Atmospheric Sciences},
  langid = {english}
}

@article{gilpinChaosInterpretableBenchmark,
  title = {Chaos as an Interpretable Benchmark for Forecasting and Data-Driven Modelling},
  author = {Gilpin, William},
  year = {2023},
  month = jan,
  journal = {arXiv preprint},
  number = {arXiv:2110.05266},
  eprint = {2110.05266},
  primaryclass = {nlin},
  publisher = {{arXiv}},
  doi = {10.48550/arXiv.2110.05266},
  urldate = {2023-10-01},
  archiveprefix = {arxiv}
}

@book{sprottChaosTimeSeriesAnalysis2001,
  title = {Chaos and {{Time-Series Analysis}}},
  author = {Sprott, J. C.},
  year = {2001},
  month = sep,
  edition = {Illustrated edition},
  publisher = {{Oxford University Press}},
  address = {{Oxford ; New York}},
  isbn = {978-0-19-850840-3},
  langid = {english}
}

@techreport{fehlbergLoworderClassicalRungeKutta1969,
  title = {Low-Order Classical {{Runge-Kutta}} Formulas with Stepsize Control and Their Application to Some Heat Transfer Problems},
  author = {Fehlberg, E.},
  year = {1969},
  month = jul,
  number = {NASA-TR-R-315},
  urldate = {2023-10-01}
}

@incollection{langfordNumericalStudiesTorus1984,
  title = {Numerical {{Studies}} of {{Torus Bifurcations}}},
  booktitle = {Numerical {{Methods}} for {{Bifurcation Problems}}: {{Proceedings}} of the {{Conference}} at the {{University}} of {{Dortmund}}, {{August}} 22{\textendash}26, 1983},
  author = {Langford, W. F.},
  editor = {K{\"u}pper, T. and Mittelmann, H. D. and Weber, H.},
  year = {1984},
  series = {International {{Series}} of {{Numerical Mathematics}}},
  pages = {285--295},
  publisher = {{Birkh{\"a}user}},
  address = {{Basel}},
  doi = {10.1007/978-3-0348-6256-1_19},
  urldate = {2024-01-28},
  isbn = {978-3-0348-6256-1},
  langid = {english}
}

@book{tarantolaInverseProblemTheory2005,
  title = {Inverse {{Problem Theory}} and {{Methods}} for {{Model Parameter Estimation}}},
  author = {Tarantola, Albert},
  year = {2005},
  month = jan,
  series = {Other {{Titles}} in {{Applied Mathematics}}},
  publisher = {{Society for Industrial and Applied Mathematics}},
  doi = {10.1137/1.9780898717921},
  urldate = {2023-11-16},
  isbn = {978-0-89871-572-9}
}

@phdthesis{harrisInferringParametricVariation2021,
  title = {Inferring Parametric Variation across Non-Stationary Time Series},
  author = {Harris, Brendan},
  year = {2021},
  langid = {english},
  school = {The University of Sydney}
}

@article{goldbergerPhysioBankPhysioToolkitPhysioNet2000,
  title = {{{PhysioBank}}, {{PhysioToolkit}}, and {{PhysioNet}}: Components of a New Research Resource for Complex Physiologic Signals},
  shorttitle = {{{PhysioBank}}, {{PhysioToolkit}}, and {{PhysioNet}}},
  author = {Goldberger, A. L. and Amaral, L. A. and Glass, L. and Hausdorff, J. M. and Ivanov, P. C. and Mark, R. G. and Mietus, J. E. and Moody, G. B. and Peng, C. K. and Stanley, H. E.},
  year = {2000},
  month = jun,
  journal = {Circulation},
  volume = {101},
  number = {23},
  pages = {E215-220},
  issn = {1524-4539},
  doi = {10.1161/01.cir.101.23.e215},
  langid = {english},
  pmid = {10851218}
}

@article{kempAnalysisSleepdependentNeuronal2000,
  title = {Analysis of a Sleep-Dependent Neuronal Feedback Loop: The Slow-Wave Microcontinuity of the {{EEG}}},
  shorttitle = {Analysis of a Sleep-Dependent Neuronal Feedback Loop},
  author = {Kemp, B. and Zwinderman, A.H. and Tuk, B. and Kamphuisen, H.A.C. and Oberye, J.J.L.},
  year = {2000},
  month = sep,
  journal = {IEEE Transactions on Biomedical Engineering},
  volume = {47},
  number = {9},
  pages = {1185--1194},
  issn = {1558-2531},
  doi = {10.1109/10.867928},
  urldate = {2024-03-06}
}

@article{berry2012aasm,
  title={The AASM manual for the scoring of sleep and associated events},
  author={Berry, Richard B and Brooks, Rita and Gamaldo, Charlene E and Harding, Susan M and Marcus, Carole and Vaughn, Bradley V and others},
  journal={Rules, Terminology and Technical Specifications, Darien, Illinois, American Academy of Sleep Medicine},
  volume={176},
  number={2012},
  pages={7},
  year={2012}
}

@article{leeInterraterReliabilitySleep2022,
  title = {Interrater Reliability of Sleep Stage Scoring: A Meta-Analysis},
  shorttitle = {Interrater Reliability of Sleep Stage Scoring},
  author = {Lee, Yun Ji and Lee, Jae Yong and Cho, Jae Hoon and Choi, Ji Ho},
  year = {2022},
  month = jan,
  journal = {Journal of Clinical Sleep Medicine : JCSM : Official Publication of the American Academy of Sleep Medicine},
  volume = {18},
  number = {1},
  pages = {193--202},
  issn = {1550-9389},
  doi = {10.5664/jcsm.9538},
  urldate = {2024-03-06},
  pmcid = {PMC8807917},
  pmid = {34310277}
}

@misc{WiskottlabSklearnsfa2024,
  title = {https://github.com/wiskott-lab/sklearn-sfa},
  year = {2024},
  month = jan,
  urldate = {2024-03-06},
  copyright = {BSD-3-Clause},
  howpublished = {wiskott-lab}
}

@article{kendallNewMeasureRank1938,
  title = {A {{New Measure}} of {{Rank Correlation}}},
  author = {Kendall, M. G.},
  year = {1938},
  journal = {Biometrika},
  volume = {30},
  number = {1/2},
  eprint = {2332226},
  eprinttype = {jstor},
  pages = {81--93},
  publisher = {[Oxford University Press, Biometrika Trust]},
  issn = {0006-3444},
  doi = {10.2307/2332226},
  urldate = {2024-03-08}
}

@article{nguyenNewInvariantMeasures2015,
  title = {New Invariant Measures to Track Slow Parameter Drifts in Fast Dynamical Systems},
  author = {Nguyen, Son Hai and Chelidze, David},
  year = {2015},
  month = jan,
  journal = {Nonlinear Dynamics},
  volume = {79},
  number = {2},
  pages = {1207--1216},
  issn = {0924-090X, 1573-269X},
  doi = {10.1007/s11071-014-1737-y},
  urldate = {2023-11-17},
  langid = {english}
}

@misc{hendersonNeverDullMoment2023,
  title = {Never a {{Dull Moment}}: {{Distributional Properties}} as a {{Baseline}} for {{Time-Series Classification}}},
  shorttitle = {Never a {{Dull Moment}}},
  author = {Henderson, Trent and Bryant, Annie G. and Fulcher, Ben D.},
  year = {2023},
  month = mar,
  number = {arXiv:2303.17809},
  eprint = {2303.17809},
  primaryclass = {cs, stat},
  publisher = {arXiv},
  doi = {10.48550/arXiv.2303.17809},
  urldate = {2023-11-22},
  archiveprefix = {arxiv}
}

@article{blasiusComplexDynamicsPhase1999,
  title = {Complex Dynamics and Phase Synchronization in Spatially Extended Ecological Systems},
  author = {Blasius, Bernd and Huppert, Amit and Stone, Lewi},
  year = {1999},
  month = may,
  journal = {Nature},
  volume = {399},
  number = {6734},
  pages = {354--359},
  publisher = {Nature Publishing Group},
  issn = {1476-4687},
  doi = {10.1038/20676},
  urldate = {2024-07-04},
  copyright = {1999 Macmillan Magazines Ltd.},
  langid = {english}
}

@article{cliffUnifyingPairwiseInteractions2023a,
  title = {Unifying Pairwise Interactions in Complex Dynamics},
  author = {Cliff, Oliver M. and Bryant, Annie G. and Lizier, Joseph T. and Tsuchiya, Naotsugu and Fulcher, Ben D.},
  year = {2023},
  month = oct,
  journal = {Nature Computational Science},
  volume = {3},
  number = {10},
  pages = {883--893},
  publisher = {Nature Publishing Group},
  issn = {2662-8457},
  doi = {10.1038/s43588-023-00519-x},
  urldate = {2024-07-10},
  copyright = {2023 The Author(s), under exclusive licence to Springer Nature America, Inc.},
  langid = {english}
}

@inproceedings{hendersonEmpiricalEvaluationTimeSeries2021,
  title = {An {{Empirical Evaluation}} of {{Time-Series Feature Sets}}},
  booktitle = {2021 {{International Conference}} on {{Data Mining Workshops}} ({{ICDMW}})},
  author = {Henderson, Trent and Fulcher, Ben D.},
  year = {2021},
  month = dec,
  pages = {1032--1038},
  issn = {2375-9259},
  doi = {10.1109/ICDMW53433.2021.00134},
  urldate = {2023-11-17}
}

@article{peachHCGAHighlyComparative2021,
  title = {{{HCGA}}: {{Highly}} Comparative Graph Analysis for Network Phenotyping},
  shorttitle = {{{HCGA}}},
  author = {Peach, Robert L. and Arnaudon, Alexis and Schmidt, Julia A. and Palasciano, Henry A. and Bernier, Nathan R. and Jelfs, Kim E. and Yaliraki, Sophia N. and Barahona, Mauricio},
  year = {2021},
  month = apr,
  journal = {Patterns},
  volume = {2},
  number = {4},
  pages = {100227},
  issn = {2666-3899},
  doi = {10.1016/j.patter.2021.100227},
  urldate = {2024-07-17}
}

@article{cabralFATSFeetsFurther2018,
  title = {From {{FATS}} to Feets: {{Further}} Improvements to an Astronomical Feature Extraction Tool Based on Machine Learning},
  shorttitle = {From {{FATS}} to Feets},
  author = {Cabral, J. B. and S{\'a}nchez, B. and Ramos, F. and Gurovich, S. and Granitto, P. M. and Vanderplas, J.},
  year = {2018},
  month = oct,
  journal = {Astronomy and Computing},
  volume = {25},
  pages = {213--220},
  issn = {2213-1337},
  doi = {10.1016/j.ascom.2018.09.005},
  urldate = {2024-07-17}
}

@article{rodgersThirteenWaysLook,
  title = {Thirteen {{Ways}} to {{Look}} at the {{Correlation Coefficient}}},
  author = {Rodgers, Joseph Lee and Nicewander, W Alan},
  journal = {The American Statistician},
  year = {1988},
  langid = {english}
}

@misc{alamCanonicalTimeseriesFeatures2024,
  title = {Canonical Time-Series Features for Characterizing Biologically Informative Dynamical Patterns in {{fMRI}}},
  author = {Alam, Imran and Harris, Brendan and Cahill, Patrick and Cliff, Oliver and Markicevic, Marija and Zerbi, Valerio and Fulcher, Ben D.},
  year = {2024},
  month = jul,
  primaryclass = {New Results},
  pages = {2024.07.14.603477},
  publisher = {bioRxiv},
  doi = {10.1101/2024.07.14.603477},
  urldate = {2024-07-18},
  archiveprefix = {bioRxiv},
  chapter = {New Results},
  copyright = {{\copyright} 2024, Posted by Cold Spring Harbor Laboratory. This pre-print is available under a Creative Commons License (Attribution 4.0 International), CC BY 4.0, as described at http://creativecommons.org/licenses/by/4.0/},
  langid = {english}
}

@article{metznerSleepRandomWalk2021,
  title = {Sleep as a Random Walk: A Super-Statistical Analysis of {{EEG}} Data across Sleep Stages},
  shorttitle = {Sleep as a Random Walk},
  author = {Metzner, Claus and Schilling, Achim and Traxdorf, Maximilian and Schulze, Holger and Krauss, Patrick},
  year = {2021},
  month = dec,
  journal = {Communications Biology},
  volume = {4},
  number = {1},
  pages = {1--11},
  publisher = {Nature Publishing Group},
  issn = {2399-3642},
  doi = {10.1038/s42003-021-02912-6},
  urldate = {2024-07-24},
  copyright = {2021 The Author(s)},
  langid = {english}
}

@article{decatTraditionalSleepScoring2022a,
  title = {Beyond Traditional Sleep Scoring: {{Massive}} Feature Extraction and Data-Driven Clustering of Sleep Time Series},
  shorttitle = {Beyond Traditional Sleep Scoring},
  author = {Decat, Nicolas and Walter, Jasmine and Koh, Zhao H. and Sribanditmongkol, Piengkwan and Fulcher, Ben D. and Windt, Jennifer M. and Andrillon, Thomas and Tsuchiya, Naotsugu},
  year = {2022},
  month = oct,
  journal = {Sleep Medicine},
  volume = {98},
  pages = {39--52},
  issn = {1878-5506},
  doi = {10.1016/j.sleep.2022.06.013},
  langid = {english},
  pmid = {35779380}
}

@article{wuAssessSleepStage2015,
  title = {Assess Sleep Stage by Modern Signal Processing Techniques},
  author = {Wu, Hau-Tieng and Talmon, Ronen and Lo, Yu-Lun},
  year = {2015},
  month = apr,
  journal = {IEEE transactions on bio-medical engineering},
  volume = {62},
  number = {4},
  pages = {1159--1168},
  issn = {1558-2531},
  doi = {10.1109/TBME.2014.2375292},
  langid = {english},
  pmid = {25438301}
}

@article{katzAlternatingDiffusionMaps2019,
  title = {Alternating Diffusion Maps for Multimodal Data Fusion},
  author = {Katz, Ori and Talmon, Ronen and Lo, Yu-Lun and Wu, Hau-Tieng},
  year = {2019},
  month = jan,
  journal = {Information Fusion},
  volume = {45},
  pages = {346--360},
  issn = {1566-2535},
  doi = {10.1016/j.inffus.2018.01.007},
  urldate = {2025-01-16}
}

@incollection{liuExploreIntrinsicGeometry2021,
  title = {Explore {{Intrinsic Geometry}} of {{Sleep Dynamics}} and {{Predict Sleep Stage}} by {{Unsupervised Learning Techniques}}},
  booktitle = {Harmonic {{Analysis}} and {{Applications}}},
  author = {Liu, Gi-Ren and Lo, Yu-Lun and Sheu, Yuan-Chung and Wu, Hau-Tieng},
  editor = {Rassias, Michael Th.},
  year = {2021},
  pages = {279--324},
  publisher = {Springer International Publishing},
  address = {Cham},
  doi = {10.1007/978-3-030-61887-2_11},
  urldate = {2025-01-16},
  isbn = {978-3-030-61887-2},
  langid = {english}
}

@article{banvilleUncoveringStructureClinical2021,
  title = {Uncovering the Structure of Clinical {{EEG}} Signals with Self-Supervised Learning},
  author = {Banville, Hubert and Chehab, Omar and Hyv{\"a}rinen, Aapo and Engemann, Denis-Alexander and Gramfort, Alexandre},
  year = {2021},
  month = mar,
  journal = {Journal of Neural Engineering},
  volume = {18},
  number = {4},
  pages = {046020},
  publisher = {IOP Publishing},
  issn = {1741-2552},
  doi = {10.1088/1741-2552/abca18},
  urldate = {2025-01-16},
  langid = {english}
}

@article{yangSelfSupervisedElectroencephalogramRepresentation2023,
  title = {Self-{{Supervised Electroencephalogram Representation Learning}} for {{Automatic Sleep Staging}}: {{Model Development}} and {{Evaluation Study}}},
  shorttitle = {Self-{{Supervised Electroencephalogram Representation Learning}} for {{Automatic Sleep Staging}}},
  author = {Yang, Chaoqi and Xiao, Cao and Westover, M. Brandon and Sun, Jimeng},
  year = {2023},
  journal = {Jmir Ai},
  volume = {2},
  number = {1},
  pages = {e46769},
  issn = {2817-1705},
  doi = {10.2196/46769},
  langid = {english},
  pmcid = {PMC10715804},
  pmid = {38090533}
}

@article{metznerExtractingContinuousSleep2023,
  title = {Extracting Continuous Sleep Depth from {{EEG}} Data without Machine Learning},
  author = {Metzner, Claus and Schilling, Achim and Traxdorf, Maximilian and Schulze, Holger and Tziridis, Konstantin and Krauss, Patrick},
  year = {2023},
  month = may,
  journal = {Neurobiology of Sleep and Circadian Rhythms},
  volume = {14},
  pages = {100097},
  issn = {2451-9944},
  doi = {10.1016/j.nbscr.2023.100097},
  langid = {english},
  pmcid = {PMC10238579},
  pmid = {37275555}
}

@article{yeCoSleepMultiViewRepresentation2022,
  title = {{{{\emph{CoSleep}}}} : {{A Multi-View Representation Learning Framework}} for {{Self-Supervised Learning}} of {{Sleep Stage Classification}}},
  shorttitle = {{{{\emph{CoSleep}}}}},
  author = {Ye, Jianan and Xiao, Qinfeng and Wang, Jing and Zhang, Hongjun and Deng, Jiaoxue and Lin, Youfang},
  year = {2022},
  journal = {IEEE Signal Processing Letters},
  volume = {29},
  pages = {189--193},
  issn = {1070-9908, 1558-2361},
  doi = {10.1109/LSP.2021.3130826},
  urldate = {2025-01-16},
  copyright = {https://ieeexplore.ieee.org/Xplorehelp/downloads/license-information/IEEE.html}
}

@article{mcphersonCharacterizationSleepUsing2001,
  title = {Characterization of Sleep Using Bispectral Analysis},
  author = {McPherson, C. and Behbehani, K. and {Dzu Dao} and Burk, J. and Lucas, E.},
  year = {2001},
  journal = {2001 Conference Proceedings of the 23rd Annual International Conference of the IEEE Engineering in Medicine and Biology Society},
  volume = {3},
  pages = {2216--2219},
  publisher = {IEEE},
  address = {Istanbul, Turkey},
  doi = {10.1109/IEMBS.2001.1017212},
  urldate = {2025-01-16},
  isbn = {9780780372115}
}

@article{buriokaApproximateEntropyElectroencephalogram2005,
  title = {Approximate {{Entropy}} in the {{Electroencephalogram}} during {{Wake}} and {{Sleep}}},
  author = {Burioka, Naoto and Miyata, Masanori and Corn{\'e}lissen, Germaine and Halberg, Franz and Takeshima, Takao and Kaplan, Daniel T. and Suyama, Hisashi and Endo, Masanori and Maegaki, Yoshihiro and Nomura, Takashi and Tomita, Yutaka and Nakashima, Kenji and Shimizu, Eiji},
  year = {2005},
  month = jan,
  journal = {Clinical EEG and Neuroscience},
  volume = {36},
  number = {1},
  pages = {21--24},
  publisher = {SAGE Publications Inc},
  issn = {1550-0594},
  doi = {10.1177/155005940503600106},
  urldate = {2025-01-17},
  langid = {english}
}

@article{patelUsingMachineLearning2021a,
  title = {Using Machine Learning to Predict Statistical Properties of Non-Stationary Dynamical Processes: {{System}} Climate,Regime Transitions, and the Effect of Stochasticity},
  shorttitle = {Using Machine Learning to Predict Statistical Properties of Non-Stationary Dynamical Processes},
  author = {Patel, Dhruvit and Canaday, Daniel and Girvan, Michelle and Pomerance, Andrew and Ott, Edward},
  year = {2021},
  month = mar,
  journal = {Chaos: An Interdisciplinary Journal of Nonlinear Science},
  volume = {31},
  number = {3},
  pages = {033149},
  issn = {1054-1500},
  doi = {10.1063/5.0042598},
  urldate = {2025-01-24}
}

@article{tokudaPredictionUnobservedBifurcation2024,
  title = {Prediction of Unobserved Bifurcation by Unsupervised Extraction of Slowly Time-Varying System Parameter Dynamics from Time Series Using Reservoir Computing},
  author = {Tokuda, Keita and Katori, Yuichi},
  year = {2024},
  month = oct,
  journal = {Frontiers in Artificial Intelligence},
  volume = {7},
  publisher = {Frontiers},
  issn = {2624-8212},
  doi = {10.3389/frai.2024.1451926},
  urldate = {2025-01-24},
  langid = {english}
}

@article{nicolaouDatadrivenDiscoveryExtrapolation2023c,
  title = {Data-Driven Discovery and Extrapolation of Parameterized Pattern-Forming Dynamics},
  author = {Nicolaou, Zachary G. and Huo, Guanyu and Chen, Yihui and Brunton, Steven L. and Kutz, J. Nathan},
  year = {2023},
  month = nov,
  journal = {Physical Review Research},
  volume = {5},
  number = {4},
  pages = {L042017},
  publisher = {American Physical Society},
  doi = {10.1103/PhysRevResearch.5.L042017},
  urldate = {2025-01-24}
}

@book{shalev-shwartzUnderstandingMachineLearning2014,
  title = {Understanding {{Machine Learning}}: {{From Theory}} to {{Algorithms}}},
  shorttitle = {Understanding {{Machine Learning}}},
  author = {{Shalev-Shwartz}, Shai and {Ben-David}, Shai},
  year = {2014},
  month = may,
  edition = {1st edition},
  publisher = {Cambridge University Press},
  address = {New York},
  isbn = {978-1-107-05713-5},
  langid = {english}
}

@misc{lubbaDynamicsAndNeuralSystemsCatch22V02022,
  title = {{{DynamicsAndNeuralSystems}}/Catch22: V0.4.0},
  shorttitle = {{{DynamicsAndNeuralSystems}}/Catch22},
  author = {Lubba, Carl H. and Fulcher, Ben and Henderson, Trent and Harris, Brendan and Cliff, Oliver and {Olivier-tl}},
  year = {2022},
  month = jun,
  doi = {10.5281/zenodo.6673597},
  urldate = {2025-02-26},
  howpublished = {Zenodo}
}

@article{virtanenSciPyFundamentalAlgorithms2020,
  title = {{{SciPy}} 1.0: Fundamental Algorithms for Scientific Computing in {{Python}}},
  shorttitle = {{{SciPy}} 1.0},
  author = {Virtanen, Pauli and Gommers, Ralf and Oliphant, Travis E. and Haberland, Matt and Reddy, Tyler and Cournapeau, David and Burovski, Evgeni and Peterson, Pearu and Weckesser, Warren and Bright, Jonathan and {van der Walt}, St{\'e}fan J. and Brett, Matthew and Wilson, Joshua and Millman, K. Jarrod and Mayorov, Nikolay and Nelson, Andrew R. J. and Jones, Eric and Kern, Robert and Larson, Eric and Carey, C. J. and Polat, {\.I}lhan and Feng, Yu and Moore, Eric W. and VanderPlas, Jake and Laxalde, Denis and Perktold, Josef and Cimrman, Robert and Henriksen, Ian and Quintero, E. A. and Harris, Charles R. and Archibald, Anne M. and Ribeiro, Ant{\^o}nio H. and Pedregosa, Fabian and {van Mulbregt}, Paul},
  year = {2020},
  month = mar,
  journal = {Nature Methods},
  volume = {17},
  number = {3},
  pages = {261--272},
  publisher = {Nature Publishing Group},
  issn = {1548-7105},
  doi = {10.1038/s41592-019-0686-2},
  urldate = {2025-02-26},
  copyright = {2020 The Author(s)},
  langid = {english}
}

@article{rollinsonWorkingSpaceTime2021,
  title = {Working across Space and Time: Nonstationarity in Ecological Research and Application},
  shorttitle = {Working across Space and Time},
  author = {Rollinson, Christine R and Finley, Andrew O and Alexander, M Ross and Banerjee, Sudipto and Dixon Hamil, Kelly-Ann and Koenig, Lauren E and Locke, Dexter Henry and DeMarche, Megan L and Tingley, Morgan W and Wheeler, Kathryn and Youngflesh, Casey and Zipkin, Elise F},
  year = {2021},
  journal = {Frontiers in Ecology and the Environment},
  volume = {19},
  number = {1},
  pages = {66--72},
  issn = {1540-9309},
  doi = {10.1002/fee.2298},
  urldate = {2025-03-19},
  copyright = {Frontiers in Ecology and the Environment{\copyright} 2020 The Authors. Frontiers in Ecology and the Environment published by Wiley Periodicals LLC on behalf of Ecological Society of America.},
  langid = {english}
}

@article{gourevitchSimpleIndicatorNonstationarity2007,
  title = {A Simple Indicator of Nonstationarity of Firing Rate in Spike Trains},
  author = {Gour{\'e}vitch, Boris and Eggermont, Jos J.},
  year = {2007},
  month = jun,
  journal = {Journal of Neuroscience Methods},
  volume = {163},
  number = {1},
  pages = {181--187},
  issn = {0165-0270},
  doi = {10.1016/j.jneumeth.2007.02.021},
  urldate = {2025-11-17}
}

@article{slaterNonstationaryWeatherWater2021,
  title = {Nonstationary Weather and Water Extremes: A Review of Methods for Their Detection, Attribution, and Management},
  shorttitle = {Nonstationary Weather and Water Extremes},
  author = {Slater, Louise J. and Anderson, Bailey and Buechel, Marcus and Dadson, Simon and Han, Shasha and Harrigan, Shaun and Kelder, Timo and Kowal, Katie and Lees, Thomas and Matthews, Tom and Murphy, Conor and Wilby, Robert L.},
  year = {2021},
  month = jul,
  journal = {Hydrology and Earth System Sciences},
  volume = {25},
  number = {7},
  pages = {3897--3935},
  publisher = {Copernicus GmbH},
  issn = {1027-5606},
  doi = {10.5194/hess-25-3897-2021},
  urldate = {2025-11-17},
  langid = {english}
}

@article{owensTimeseriesDimensionReduction,
  title = {Time-Series Dimension Reduction: A Comprehensive Review and Conceptual Unification of Algorithms},
  shorttitle = {Time-Series Dimension Reduction},
  author = {Owens, Kieran S. and Fulcher, Ben D.},
  urldate = {2025-12-03},
  publisher = {{TechRxiv}},
  doi = {https://doi.org/10.36227/techrxiv.176409662.29359239/v1},
  year = {2025},
  journal = {TechRxiv}
}

@article{owensParameterInferenceNonstationary2024,
  title = {Parameter Inference from a Non-Stationary Unknown Process},
  author = {Owens, Kieran S. and Fulcher, Ben D.},
  year = {2024},
  journal = {Chaos: An Interdisciplinary Journal of Nonlinear Science},
  volume = {34},
  number = {10},
  doi = {10.1063/5.0228236}
}

@book{hyvarinenIndependentComponentAnalysis2001,
  title = {Independent {{Component Analysis}}: 26},
  shorttitle = {Independent {{Component Analysis}}},
  author = {Hyv{\"a}rinen, Aapo and Karhunen, Juha and Oja, Erkki},
  year = {2001},
  month = may,
  edition = {1st edition},
  publisher = {Wiley-Interscience},
  address = {New York},
  isbn = {978-0-471-40540-5},
  langid = {english}
}

@article{hendersonFeatureBasedTimeSeriesAnalysis2025,
  title = {Feature-{{Based Time-Series Analysis}} in {{R}} Using the {{Theft Ecosystem}}},
  author = {Henderson, Trent and Fulcher, Ben D.},
  year = {2025},
  month = oct,
  journal = {The R Journal},
  volume = {17},
  number = {3},
  pages = {43--68},
  issn = {2073-4859},
  doi = {10.32614/RJ-2025-023},
  urldate = {2026-05-25}
}

@article{uriguenEEGArtifactRemoval2015a,
  title = {{{EEG}} Artifact Removal---State-of-the-Art and Guidelines},
  author = {Urig{\"u}en, Jose Antonio and {Garcia-Zapirain}, Bego{\~n}a},
  year = {2015},
  month = apr,
  journal = {Journal of Neural Engineering},
  volume = {12},
  number = {3},
  pages = {031001},
  publisher = {IOP Publishing},
  issn = {1741-2552},
  doi = {10.1088/1741-2560/12/3/031001},
  urldate = {2026-05-27},
  langid = {english}
}

@article{moorePyhctsaPythonPackage2026,
  title = {Pyhctsa: {{A Python}} Package for Highly Comparative Time-Series Analysis},
  shorttitle = {Pyhctsa},
  author = {Moore, Joshua B. and Fulcher, Ben D.},
  year = {2026},
  month = jul,
  journal = {Journal of Open Source Software},
  volume = {11},
  number = {123},
  pages = {10581},
  issn = {2475-9066},
  doi = {10.21105/joss.10581},
  urldate = {2026-07-20},
  langid = {english}
}

@misc{fulcherInterpretableModelfreeInference2026,
  title = {Interpretable Model-Free Inference of Parametric Variation across Time-Series Data through Large-Scale Feature Extraction},
  author = {Fulcher, Ben D. and Lubba, Carl H. and Gilestro, Giorgio F. and Schultz, Simon R. and Jones, Nick S.},
  year = {2026},
  month = jun,
  journal = {arXiv.org},
  urldate = {2026-07-27},
  howpublished = {https://arxiv.org/abs/2606.12836v1},
  langid = {english}
}

@article{scottFeasibilityMultivariateDensity1991,
  title = {Feasibility of Multivariate Density Estimates},
  author = {Scott, David W.},
  year = {1991},
  month = mar,
  journal = {Biometrika},
  volume = {78},
  number = {1},
  pages = {197--205},
  issn = {0006-3444},
  doi = {10.1093/biomet/78.1.197},
  urldate = {2026-08-20}
}

@misc{otneimNonParametricEstimationConditional2016,
  type = {{{SSRN Scholarly Paper}}},
  title = {Non-{{Parametric Estimation}} of {{Conditional Densities}}: {{A New Method}}},
  shorttitle = {Non-{{Parametric Estimation}} of {{Conditional Densities}}},
  author = {Otneim, H{\aa}kon and Tjostheim, Dag},
  year = {2016},
  month = dec,
  number = {2882022},
  address = {Rochester, NY},
  doi = {10.2139/ssrn.2882022},
  urldate = {2026-08-20},
  langid = {english}
}

@article{tamakiSurveillanceREMSleep2019,
  title = {Surveillance {{During REM Sleep}} for the {{First-Night Effect}}},
  author = {Tamaki, Masako and Sasaki, Yuka},
  year = {2019},
  month = oct,
  journal = {Frontiers in Neuroscience},
  volume = {13},
  publisher = {Frontiers},
  issn = {1662-453X},
  doi = {10.3389/fnins.2019.01161},
  urldate = {2026-08-27},
  langid = {english}
}

@article{rechtschaffenAuditoryAwakeningThresholds1966,
  title = {Auditory {{Awakening Thresholds}} in {{Rem}} and {{Nrem Sleep Stages}}},
  author = {Rechtschaffen, Allan and Hauri, Peter and Zeitlin, Maurice},
  year = {1966},
  month = jun,
  journal = {Perceptual and Motor Skills},
  volume = {22},
  number = {3},
  pages = {927--942},
  publisher = {SAGE Publications Inc},
  issn = {0031-5125},
  doi = {10.2466/pms.1966.22.3.927},
  urldate = {2026-08-27},
  langid = {english}
}

@article{bonnetThresholdSleepPerception1982,
  title = {The Threshold of Sleep: Perception of Sleep as a Function of Time Asleep and Auditory Threshold},
  shorttitle = {The Threshold of Sleep},
  author = {Bonnet, M. H. and Moore, S. E.},
  year = {1982},
  journal = {Sleep},
  volume = {5},
  number = {3},
  pages = {267--276},
  issn = {0161-8105},
  doi = {10.1093/sleep/5.3.267},
  langid = {english},
  pmid = {7134732}
}

\end{document}